\documentclass[fleqn,usenatbib]{mnras}

\usepackage{newtxtext,newtxmath}

\usepackage[T1]{fontenc}

\DeclareRobustCommand{\VAN}[3]{#2}
\let\VANthebibliography\thebibliography
\def\thebibliography{\DeclareRobustCommand{\VAN}[3]{##3}\VANthebibliography}

\usepackage{graphicx}	
\usepackage{amsmath}	
\usepackage{soul}
\usepackage{xcolor}
\definecolor{my_blue}{HTML}{91C4E8}
\definecolor{my_green}{HTML}{AEDCA8}
\definecolor{my_red}{HTML}{F29D9D}
\definecolor{my_purple}{HTML}{CBB1D8}

\title[Gas accretion in Auriga Milky Way-like galaxies]{Cosmological gas accretion history onto the stellar discs of Milky Way-like galaxies in the Auriga simulations -- (II) The inside-out growth of discs 
}

\author[F.~G. Iza et al.]{Federico~G.~Iza$^{1,2}$\thanks{E-mail: fiza@iafe.uba.ar},
Sebasti\'an~E.~Nuza$^{1}$,
Cecilia~Scannapieco$^{2}$,
Robert~J.~J.~Grand$^{3,4}$, \newauthor
Facundo~A.~Gómez$^{5,6}$,
Volker~Springel$^{7}$,
Rüdiger~Pakmor$^{7}$,
Federico Marinacci$^{8,9}$,
Francesca Fragkoudi$^{10}$
\\
\\
$^1$ Instituto de Astronom\'ia y F\'isica del Espacio (IAFE, CONICET-UBA), 1428 Buenos Aires, Argentina \\
$^2$ Facultad de Ciencias Exactas y Naturales (FCEyN), Universidad de Buenos Aires (UBA), 1428 Buenos Aires, Argentina \\
$^3$ Instituto de Astrof\'isica de Canarias, Calle V\'ia L\'actea s/n, E-38205 La Laguna, Tenerife, Spain \\
$^4$ Departamento de Astrofísica, Universidad de La Laguna, Av. del Astrofísico Francisco Sánchez s/n, E-38206 La Laguna, Tenerife, Spain \\
$^5$ Departamento de Astronom\'ia, Universidad de La Serena, Avenida Juan Cisternas 1200, La Serena, Chile \\
$^6$ Instituto de Investigaci\'on Multidisciplinar en Ciencia y Tecnolog\'ia, Universidad de La Serena, Ra\'ul Bitr\'an 1305, La Serena, Chile \\
$^7$ Max-Planck-Institut für Astrophysik, Karl-Schwarzschild-Str 1, D-85748 Garching, Germany \\
$^8$ Department of Physics \& Astronomy ``Augusto Righi'', University of Bologna, via Gobetti 93/2, I-40129 Bologna, Italy \\
$^9$ INAF, Astrophysics and Space Science Observatory Bologna, Via P. Gobetti 93/3, I-40129 Bologna, Italy \\
$^{10}$ Institute for Computational Cosmology, Department of Physics, Durham University, South Road, Durham DH1 3LE, United Kingdom}

\date{Accepted XXX. Received YYY; in original form ZZZ}

\pubyear{2023}

\begin{document}
\label{firstpage}
\pagerange{\pageref{firstpage}--\pageref{lastpage}}
\maketitle

\begin{abstract}  
\noindent We investigate the growth of stellar discs in Milky Way-mass galaxies using the magnetohydrodynamical simulations of the Auriga Project in a full cosmological context.
We focus on the gas accretion process along the discs, calculating the net, infall and outflow rates as a function of galactocentric distance, and investigate the relation between them and the star formation activity.
The stellar distributions of around $70\%$ of the simulated galaxies exhibit an ``inside-out'' pattern, with older (younger) stellar populations preferentially located in the inner (outer) disc regions.
In all cases, we find a very tight correlation between the infall, outflow and net accretion rates, as well as between these three quantities and the star formation rate.
This is because the amount of gas which is ultimately available for star formation in each radial ring depends not only on the infall rates, but also on the amount of gas leaving the disc in outflows, which directly relates to the local star formation level.
Therefore, any of these rates can be used to identify galaxies with inside-out growth.
For these galaxies, the correlation between the dominant times of accretion/star formation and disc radius is well fitted by a linear function.
We also find that, when averaged over galaxies with formation histories similar to the Milky Way, the simulated accretion rates show a similar evolution (both temporally- and radially-integrated) to the usual accretion prescriptions used in chemical evolution models, although some major differences arise at early times and in the inner disc regions.
\end{abstract}




\section{Introduction}
\label{sec:introduction}

The physical processes that lead to the formation and govern the evolution of galaxies have long been one of the main quests in modern astrophysics.
In particular, significant attention has been given to understand the formation of the Milky Way (MW) and similar spirals, as it is for the MW that we have access to the most detailed observations.
The star formation history is one of the key processes that influence the final properties of galaxies, and results from the complex interplay of the various physical processes which occur during galaxy evolution.
In this context, the variety of observed morphologies is a natural result of galaxy formation with different merger and accretion histories, resulting into a combination of disc and bulge components with varying relative significance.
While the formation of bulges is thought to occur early and in short time-scales, the discs are expected to form over longer periods, fed by gas accretion at increasingly large distances.

In this context, the stellar discs of spiral galaxies are thought to grow from the ``inside-out'' \citep[][]{Larson1976, Fall1980}, as old (young) stars preferentially populate the inner (outer) regions of galaxy discs.
This is suggested by a number of observations (see, e.g. \citealt{Frankel2019, Prantzos2023}), although inside-out growth might also depend on galaxy mass \citep{Pan2015}.
Most observations, however, rely on the present-day positions of stars instead of their unknown birth sites, but stars are subject to orbit mixing processes, therefore losing memory on their initial location.
Unlike dynamical memory, stars retain their intrinsic chemical patterns during their whole evolution, thus providing information on the properties of stars at their birth time.
As a result, the measurement of detailed chemical abundances of stars in the MW and other galaxies has opened up the possibility to investigate in detail the formation of the stellar components of spiral galaxies, especially their discs, and compare with the observed chemical abundances.

In particular, chemical evolution models (CEMs) of the MW show that observed galaxy properties can be reproduced, but the predicted stellar metallicity gradients in the discs are directly affected by the assumptions made on gas accretion rates as a function of time and radius within the disc, as gaseous material will ultimately give rise to the formation of new stellar populations \citep[e.g.][]{Matteucci12}.
In order to reproduce observational results, a first, rapid episode of gas accretion is required in CEMs to explain the formation of the spheroidal components of the Galaxy.
In contrast, accretion onto the thin-disc component is required to occur at decreasing rates during longer timescales, which is usually modelled assuming cosmic time-dependent exponential or Gaussian functions \citep[see e.g.][and references therein]{Iza22}.
Moreover, to account for the fact that less enriched (i.e. older) stars are formed towards the inner regions of the Galaxy, these models also assume that gas accretion timescales correlate with radius, thus delaying the onset of star formation in the disc outskirts \citep[e.g][]{Chiappini01}.
While this inside-out formation assumption for the disc is key in these models, further ingredients such as radial gas flows and/or variable efficiency of star formation might also play a role (see, e.g. \citealt{Palla2020}).
The combination of various assumptions is then needed in order to successfully reproduce the observed distribution of present-day chemical abundances in the Galactic disc and the photometric properties of galaxies similar to the MW \citep{Boissier99}.
However, the validity of these assumptions has not always been tested using more realistic simulations in a cosmological context that model gas flows into and out of galaxies self-consistently over their whole evolution.

Together with advances in observational studies and chemical evolution models, numerical simulations have made significant progress during the last decades, being able to reproduce the formation of galaxy discs from cosmological initial conditions (e.g. \citealt{Okamoto2005}, \citealt{Scannapieco2008}, \citealt{Guedes2011}, \citealt{Stinson2013}, \citealt{Aumer2013}, \citealt{Marinacci2014}, \citealt{Wang2015}, \citealt{Hopkins2018}).
In combination with detailed modelling of the chemical evolution of galaxies, these are a powerful tool to investigate the relation between star formation, gas accretion and morphology across cosmic time, and to study in more detail the formation of stellar discs.
In cosmological simulations, relations between gas accretion and morphology have already been found, for instance, by \cite{Aumer2014}.
As the gas component is the fuel for star formation, the star formation rate (SFR) of a galaxy will be naturally linked to its accretion history, even though processes such as feedback can make this relation somewhat complex.
In any case, if discs form from the inside-out, one might expect that gas accretion has an inside-out behaviour as well.
Recent simulation studies suggest that discs, at least at the MW scale, form from the inside-out, even though not all of them necessarily follow this pattern (e.g. \citealt{Sommer-Larsen2003, Scannapieco2009, Aumer2014, Gomez2017, Nuza2019}).

In \cite{Iza22}, the first paper of this series (Paper I), we investigated the temporal evolution of gas accretion onto the discs of simulated MW-mass galaxies.
For this, we used the original simulations of the Auriga Project \citep{Grand2017}, which consist on a sample of 30 zoom-in, MW-mass galaxy haloes simulated with the magnetohydrodynamic code {\sc arepo} \citep{Springel2010} that were selected to be relatively isolated at the present day.
The sample was separated in two different galaxy populations characterised by the behaviour of their late-time net gas accretion, with the vast majority (group G1) displaying a gentle, decaying exponential-like behaviour consistent with the expectation of CEMs, and the remaining (group G2) an increasing net accretion related to a more active merger history at late times which lead to the partial/total destruction of their discs, even though most of them ended up with new well-formed disc-like components at $z=0$.
We also found a strong correlation between the gas inflow onto the disc region and the SFR, especially at early times, confirming the close relationship between both quantities.
Furthermore, outflows also showed a strong correlation with SFR, highlighting the link between SFR and the generation of outflows following star formation, despite the fact that these processes are related in a non-trivial manner, affecting the circulation of gas in the disc-halo interface.

In this work, we focus on the evolution and growth of stellar discs as a function of radius and evaluate whether the simulations are consistent with an inside-out formation scenario by analysing both the distribution of stellar ages along the plane of the disc as well as the radial dependency of the gas accretion rates.
We quantify the inside-out behaviour of each galaxy through an {\it inside-out parameter} which provides information on the dominant times of accretion for each radial ring and can be computed for the infall, outflow, net accretion and star formation rates.
Since accretion laws are a fundamental ingredient in CEMs that aim to describe the distribution of metals in galaxies, we also compare the gas accretion results obtained from simulations with the usual assumptions adopted in CEMs.

The organisation of this paper is as follows: in Section \ref{sec:simulations_and_methods} we briefly revisit the Auriga simulations, and discuss the methods used in the analysis of the ``inside-out'' behaviour; in Section \ref{sec:results} we focus on our main results: the distribution of stellar ages along the disc, the radial dependency of the inflow, outflow and net gas accretion rates and the star formation rate (SFR), and present the average behaviour over sub-samples of the Auriga galaxies, and in Section \ref{sec:cem_comparison} we show a simple comparison of the obtained accretion rates, integrated over time and radius, with the expected accretion law from CEMs adopting the parameters of our simulated MW-like galaxy sample.
Finally, in Section~\ref{sec:conclusions}, we summarise and discuss our main results.

\begin{table*}
	\centering
	\caption{
    	Galactic properties at $z=0$.
    	The columns are:
            (1) galaxy name,
            (2) virial radius $R_{200}$,
            (3) virial mass $M_{200}$,
            (4) subhalo stellar mass $M_\star$,
            (5) subhalo gas mass $M_\mathrm{gas}$,
            (6) disc-to-total mass fraction,
            (7) disc radius $R_\mathrm{d}$ and
            (8) disc height $h_\mathrm{d}$.
    	The symbol $\dagger$ in the first column identifies galaxies that include a treatment for stochastic tracer particles.
    }
	\label{tab:galactic_properties}
	\begin{tabular}{lcccccccc}
		\hline
		Galaxy   & $R_{200}$ & $M_{200}$ & $M_\star$ & $M_\mathrm{gas}$ & D/T & $R_\mathrm{d}$ & $h_\mathrm{d}$ \\
                     & [kpc] & [$10^{10}~\mathrm{M}_\odot$] & [$10^{10}~\mathrm{M}_\odot$] & [$10^{10}~\mathrm{M}_\odot$] & & [kpc] & [kpc] \\
		\hline
		Au2               & 261.7 & 191.4 & 9.4   & 12.5  & 0.81  & 33.7  & 3.1   \\
		Au3               & 239.0 & 145.8 & 8.7   & 9.7   & 0.74  & 23.6  & 2.5   \\
		Au4               & 236.3 & 140.9 & 8.8   & 12.8  & 0.38  & 21.8  & 3.6   \\
		Au6$^\dagger$     & 212.8 & 102.9 & 5.6   & 6.4   & 0.79  & 18.2  & 2.4   \\
		Au7               & 218.9 & 112.0 & 5.9   & 11.6  & 0.60  & 21.7  & 3.2   \\
		Au8               & 216.3 & 108.1 & 4.0   & 9.5   & 0.84  & 29.2  & 3.5   \\
		Au9$^\dagger$     & 215.8 & 107.3 & 7.5   & 6.5   & 0.72  & 9.9   & 1.9   \\
		Au10              & 214.0 & 104.7 & 6.2   & 8.6   & 0.73  & 8.4	  & 1.6   \\
		Au11              & 249.0 & 164.9 & 6.4   & 8.8   & 0.62  & 19.3  & 2.5   \\
		Au12              & 217.1 & 109.3 & 6.6   & 8.8   & 0.68  & 14.9  & 2.6   \\
		Au13$^\dagger$    & 223.2 & 118.8 & 6.3   & 10.6  & 0.84  & 13.0  & 2.6   \\
		Au14              & 249.4 & 165.7 & 11.7  & 14.1  & 0.61  & 17.7  & 2.8   \\
		Au15              & 225.4 & 122.2 & 4.3   & 9.3   & 0.69  & 19.2  & 3.0   \\
		Au16              & 241.4 & 150.3 & 7.0   & 10.4  & 0.88  & 31.1  & 3.1   \\
		Au17$^\dagger$    & 215.7 & 107.1 & 9.0   & 7.6   & 0.76  & 15.0  & 2.0   \\
		Au18              & 225.3 & 122.1 & 8.4   & 7.1   & 0.80  & 14.0  & 2.0   \\
		Au20              & 227.0 & 124.9 & 5.6   & 14.2  & 0.72  & 24.1  & 3.3   \\
		Au21              & 238.6 & 145.1 & 8.7   & 11.8  & 0.80  & 18.6  & 2.9   \\
		Au22              & 205.5 & 92.6  & 6.2   & 3.6   & 0.69  & 7.9	  & 1.6   \\
		Au23$^\dagger$    & 245.2 & 157.3 & 10.1  & 9.4   & 0.84  & 19.8  & 2.4   \\
		Au24$^\dagger$    & 242.7 & 152.5 & 9.3   & 9.7   & 0.74  & 24.1  & 2.8   \\
		Au25              & 225.3 & 122.1 & 3.7   & 7.9   & 0.86  & 22.8  & 2.9   \\
		Au26$^\dagger$    & 242.7 & 152.6 & 10.9  & 10.5  & 0.71  & 14.6  & 2.1   \\
		Au27              & 253.8 & 174.5 & 10.3  & 12.5  & 0.71  & 17.2  & 2.4   \\
		\hline
	\end{tabular}
\end{table*}

\begin{figure*}
    \begin{tabular}{c}
        \includegraphics[scale=.7]{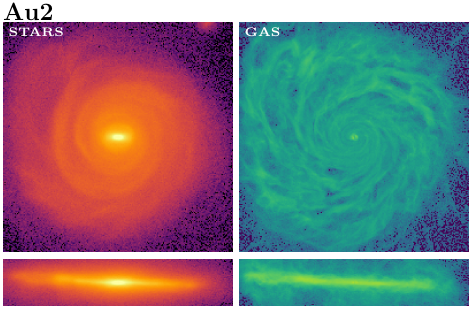}
        \includegraphics[scale=.7]{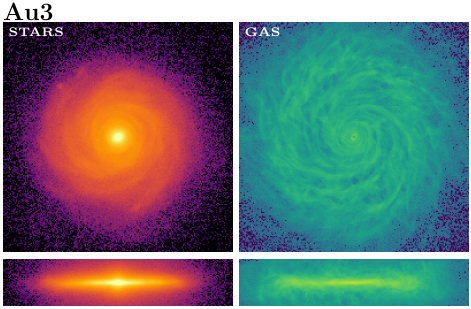}
        \includegraphics[scale=.7]{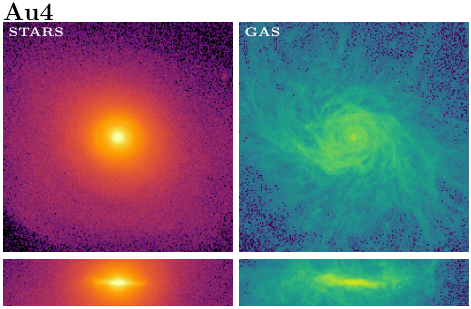} \\
        \includegraphics[scale=.7]{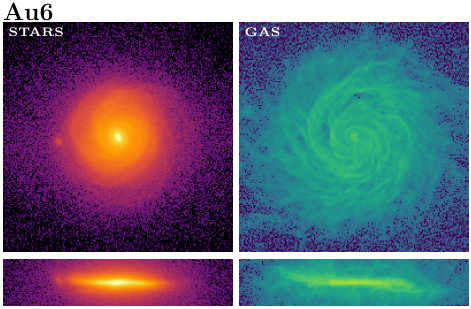}
        \includegraphics[scale=.7]{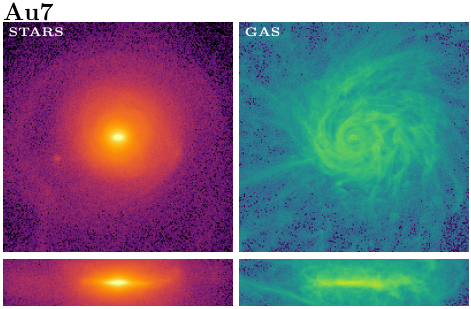}
        \includegraphics[scale=.7]{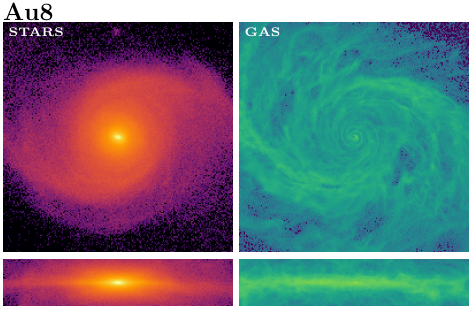} \\
        \includegraphics[scale=.7]{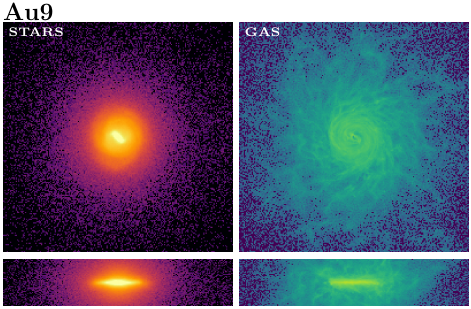}
        \includegraphics[scale=.7]{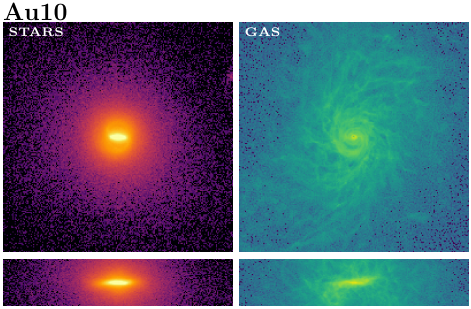}
        \includegraphics[scale=.7]{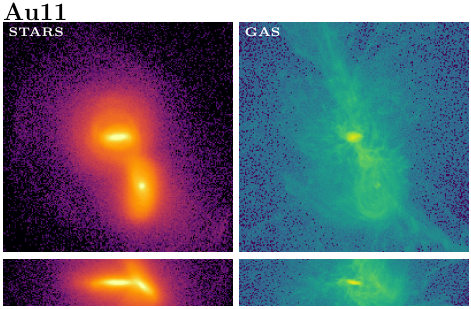} \\
        \includegraphics[scale=.7]{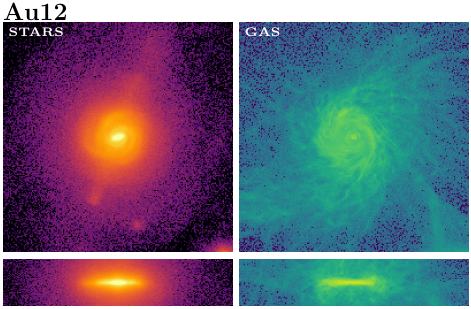}
        \includegraphics[scale=.7]{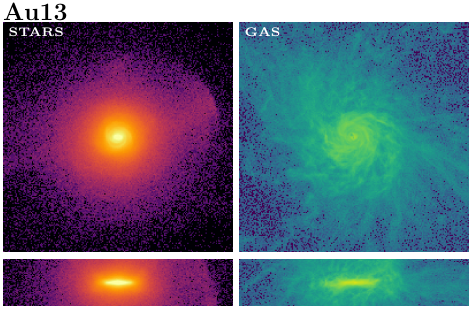}
        \includegraphics[scale=.7]{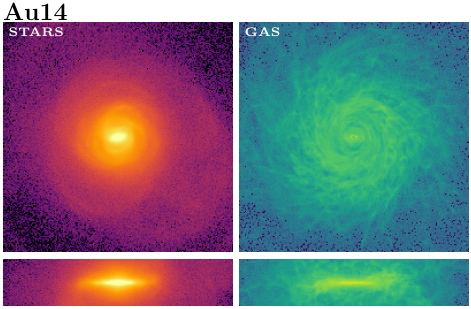}
    \end{tabular}
    \caption{
        Projected density maps for the Auriga galaxies at $z=0$, for the stellar (left) and gaseous (right) components.
        For each galaxy we show the face-on view (top) and edge-on view (bottom).
        The colour maps span five orders of magnitude in projected density using a logarithmic scale.
        The face-on view shows a cubic region of $100~\mathrm{ckpc}$ on a side and the edge-on view a region of dimensions $100 \times 100 \times 20~\mathrm{ckpc}^3$.
    }
    \label{fig:density_maps_with_gas_1}
\end{figure*}

\begin{figure*}
    \begin{tabular}{c}
        \includegraphics[scale=.7]{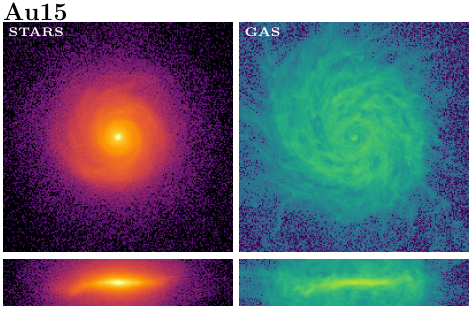}
        \includegraphics[scale=.7]{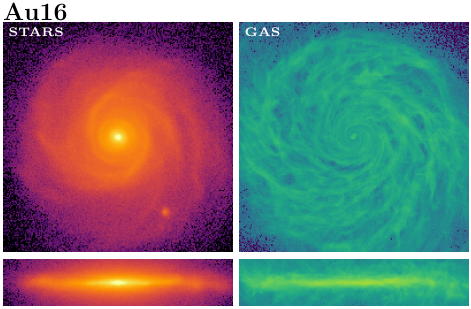}
        \includegraphics[scale=.7]{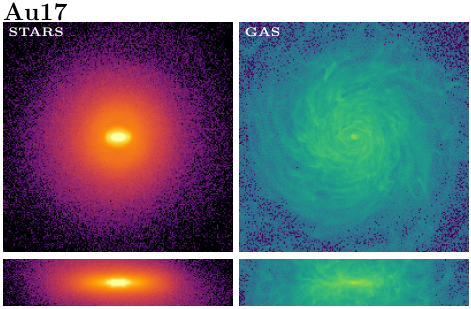} \\
        \includegraphics[scale=.7]{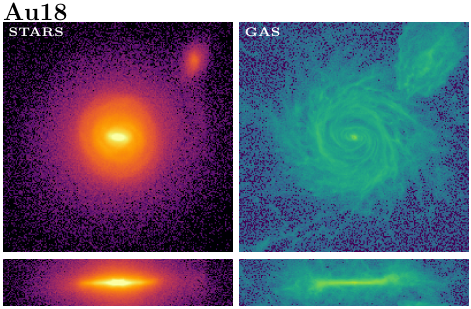}
        \includegraphics[scale=.7]{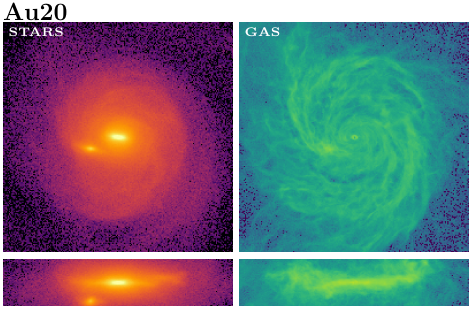}
        \includegraphics[scale=.7]{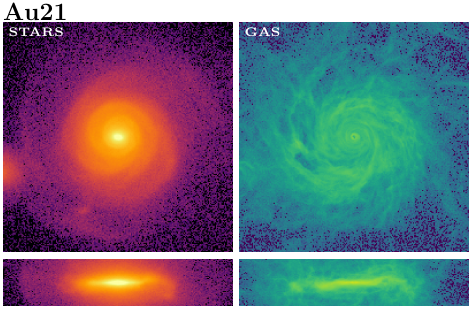} \\
        \includegraphics[scale=.7]{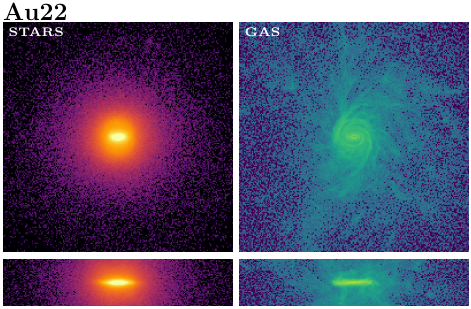}
        \includegraphics[scale=.7]{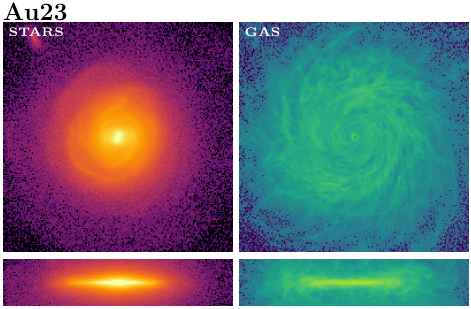}
        \includegraphics[scale=.7]{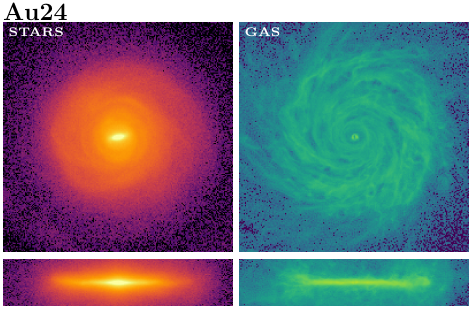} \\
        \includegraphics[scale=.7]{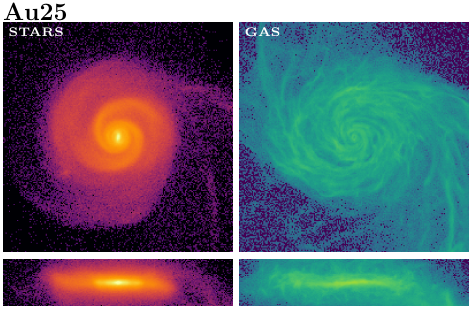}
        \includegraphics[scale=.7]{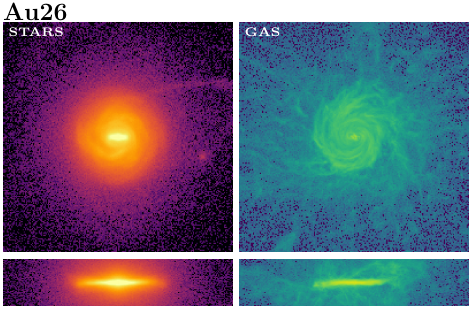}
        \includegraphics[scale=.7]{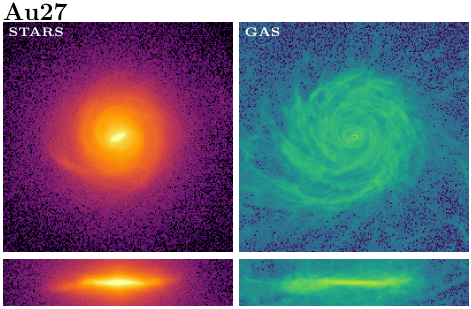}
    \end{tabular}
    \caption{
        As Fig.~\ref{fig:density_maps_with_gas_1} for the rest of the Auriga galaxies.
    }
    \label{fig:density_maps_with_gas_2}
\end{figure*}

\section{Simulations and analysis methods}
\label{sec:simulations_and_methods}

\subsection{The Auriga Simulations}
\label{sec:simulations}

The Auriga simulations \citep{Grand2017} consist on a suite of 30 Milky Way-mass disc galaxies simulated in a full cosmological setting at high resolution using the zoom-in technique.
The simulations were run with the quasi-Lagrangian, moving-mesh magnetohydrodynamic (MHD) code \textsc{arepo} \citep{Springel2010}.
Gravitational forces in \textsc{arepo} are computed with a standard TreePM method, employing a fast Fourier Transform (FFT) technique and an oct-tree algorithm for long- and short-range forces, respectively.
The MHD equations for the collisional component are discretized on a dynamic unstructured Voronoi mesh.
Further details about the implementation and usage of the \textsc{arepo} code can be found in \cite{Springel2010}.

The physical model adopted in the Auriga simulations includes a variety of physical processes such as metal-line cooling, a uniform ultraviolet background field for reionization, star formation based on the \cite{Springel2003} model, magnetic fields \citep{Pakmor2014, Pakmor2017, Pakmor2018}, energetic and chemical feedback from Type II supernovae, metal return from Type Ia supernovae and asymptotic giant branch stars and AGN feedback \citep{Vogelsberger2013, Marinacci2014, Grand2017}.

For this study, we use the level 4 resolution runs of \cite{Grand2017} with a mass resolution of $\sim 3\times 10^5 ~\mathrm{M}_\odot$ and $\sim 5\times 10^4 ~\mathrm{M}_\odot$ for dark matter and baryons, respectively.
The adopted cosmological parameters are in agreement with the estimations of \cite{Planck2014}, namely: $\Omega_{\rm M} = 0.307$, $\Omega_{\rm b} = 0.048$, $\Omega_\Lambda = 0.693$ and a Hubble constant of $H_0 = 100\,h~\mathrm{km}\,\mathrm{s}^{-1}\,\mathrm{Mpc}^{-1}$, where $h = 0.6777$.

The 30 Auriga galaxies have been selected for re-simulation at $z=0$ from a dark matter-only simulation \citep{Schaye2015}, with the conditions of having virial masses in the range $1$--$2 \times 10^{12} \, \mathrm{M}_\odot$, and being relatively isolated at the present time, with no other halo with a mass higher than $3\%$ of its own mass located within 9 times its virial radius.
From all haloes satisfying these conditions, the final Auriga sample was randomly selected from the most isolated quartile.
Further details on the Auriga simulations can be found in \cite{Grand2017}.

\subsection{Selection and properties of the Auriga galaxies}
\label{sec:auriga}

In this work, we extend the analysis of gas accretion rates on the discs presented in Paper I, where we focused on the temporal dependencies of the infall, outflow and net rates.
Here, we investigate the corresponding radial dependencies, with focus on the spatial build-up of gaseous and stellar discs and the possibility that some or most of the Auriga discs formed from the {\it inside-out} in agreement with usual assumptions for the MW.

In Paper I, we separated the Auriga galaxies in two groups according to the behaviour of the net accretion rates: galaxies with declining (constant/increasing) late-time accretion rates were classified as part of group G1 (G2).
Furthermore, G1 galaxies were identified as MW-analogues, having quiet merger histories, while the majority of galaxies in G2 experience significant merger events after the initial formation period.
In both groups, however, galaxies ended up as disc-dominated galaxies at the present.
As part of our classification, we also identified $6$ galaxies with strongly perturbed stellar discs that could affect our analysis methods and were excluded: Au1, Au5, Au19, Au28, Au29 and Au30.
For consistency with Paper I, we exclude these galaxies in the present work as well.
As a result, we are left with a total of $24$ ($19$ in G1 and $5$ in G2) galaxies, with virial masses in the range $\sim 1$--$1.7~\times10^{12} \, \mathrm{M}_\odot$ and stellar masses in the range $\sim 4$--$12~\times10^{10} \, \mathrm{M}_\odot$.

Table~\ref{tab:galactic_properties} lists the main properties of the galaxies studied in this work.
We note that all of them have virial masses consistent with the commonly accepted mass of $\sim10^{12}~\mathrm{M}_\odot$ for the MW and can be thought of as ``Milky Way-like'' galaxies in terms of their mass but they have been chosen to be relatively isolated at $z=0$, unlike in the real environment of our Galaxy.
The table also provides information on other $z=0$ properties, including an estimation of the radius and height of the galactic discs (see Section~\ref{subsec:the_stellar_disc} and Paper I for details on these calculations).

An important aspect of these simulations is that, owing to the quasi-Lagrangian nature of \textsc{arepo}, it is not possible to trace the trajectories of gas elements through cosmic time, and thus relevant information for our analysis of gas accretion is lost.
However, $7$ of the selected galaxies have been re-simulated including a treatment for stochastic tracer particles \citep{Genel2013, DeFelippis2017}, making the study of the history of gas elements possible.
These galaxies are identified with the symbol $\dagger$ in Table~\ref{tab:galactic_properties}.
As in a standard Lagrangian scheme, tracer particles can be tracked back through cosmic time allowing us to assess the evolution of gas accretion.
Further details on the simulations with tracer particles can be found in \cite{Grand2019} and \cite{Fragkoudi2021}.

\subsection{Methods}
\label{sec:methods}

\subsubsection{Defining the stellar disc}
\label{subsec:the_stellar_disc}

Computing gas accretion rates onto the discs of the Auriga galaxies as a function of cosmic time requires a consistent identification of the disc region in the simulations and a proper evaluation of its morphological properties, not only in the present, but throughout its whole evolution.
In this section, we outline the procedure used to define the disc region where inflow and outflow patterns are measured.

We first calculate, for each galaxy at a given cosmic time, the inertia tensor of all stars in the inner $10~\mathrm{ckpc}$ and take the principal axis of inertia to coincide with the $z$-axis, in a way that the latter also corresponds to the direction of the angular momentum vector of stars.
Thus, after properly rotating the reference frame of the galaxy, the galactic disc is contained in the $xy$ plane.

In Paper I we introduced a simple definition to calculate disc sizes that includes two parameters: a disc radius, $R_\mathrm{d}$, and a disc height, $h_\mathrm{d}$.
The first is defined, at each time, as the radius that encloses $90\%$ of the total host stellar mass.
In a similar fashion, the disc height for the positive ($z>0$) and negative ($z<0$) regions in the $z$-direction above and below the mid plane is computed as the distance that encloses $90\%$ of the stellar mass and we consider the disc height as the mean of both values.
In this way, $h_{\rm d}$ refers to the height above and below the disc plane resulting in a disc width $2 h_{\rm d}$.
For more details on these calculations we refer the reader to our Paper I.

This method works well for all galaxies at all times, except for very early epochs when discs have not formed yet and strong asymmetries are present in the stellar mass distribution.
We include, therefore, some corrections for cosmic time less than $4\,\mathrm{Gyr}$ together with the changes shown in disc sizes and accretion rates when considering other stellar mass fractions to define the disc (see Paper I for details).
Figures~\ref{fig:density_maps_with_gas_1} and \ref{fig:density_maps_with_gas_2} show the face-on and edge-on stellar and gaseous density maps for the simulated galaxies.

\subsubsection{Calculation of infall, outflow and net accretion rates}
\label{sec:inflow_outflow_rates_calculation}

In this section, we describe the method used to quantify the accretion rates onto the galactic discs.
To properly consider differences between galactic disc sizes and their time evolution, we divide each disc region in ten radial bins of width $0.1 R_\mathrm{d}$ and height $h_\mathrm{d}$.
Both of these parameters were calculated in Paper I for all simulations to which we refer the reader for further details.

In the simulations that do contain tracer particles, we compute, for each radial ring, the number of tracer particles that were previously outside the stellar disc that are later found inside in the next snapshot.
Since each tracer is considered to carry a fixed amount of mass, the total infalling mass between snapshots corresponds to the number of infalling tracers multiplied by the tracer mass.
Outflowing mass, on the other hand, is calculated considering the amount of tracers that were previously inside a given ring that are later found outside the stellar disc.
To compute the corresponding gas rates, the inflowing/outflowing gas masses are divided by the time span of the two considered snapshots.

Note that the inflow and outflow rates do not include radial mass flows.
Including radial flows has, by definition, the effect of increasing the inflow/outflow rates, but does not produce any  important variation to the results obtained in this work, particularly with our findings related to the inside-out growth of the stellar discs.
An analysis of radial flows of gas inside the discs of the Auriga galaxies can be found in \cite{Okalidis2021}.

The net accretion rate, on the other hand, is related to the change of gas mass inside each ring, which is, in turn, affected by both infalling/outflowing gas particles and star formation, being the latter responsible of gas depletion within the disc.
To quantify $\dot{M}_\mathrm{net}$ we use the formula detailed in Paper I, which uses information of the cells and can, therefore, be applied to all the galaxies in the sample.
Then, for each radial ring, the net gas accretion rate is given by
\begin{equation}
    \dot{M}_\mathrm{net} = \frac{M_\mathrm{gas}(i) - M_\mathrm{gas}(i-1) + M_\star}{t(i) - t(i-1)},
\end{equation}
where $M_\mathrm{gas}$ is the gas mass in the stellar disc, $i$ and $i-1$ are indexes labelling consecutive snapshots, $M_\star$ is the amount of stellar mass formed between them and $t$ is the cosmic time.
This formula accounts for the loss of gas mass in the stellar disc owing to feedback processes removing material from the disc and star formation.
Furthermore, in Paper I we showed that this  way of calculating the net accretion rates when tracer particles are not present yields similar results when compared to the ``exact'', particle-by-particle, method.

\begin{figure*}
    \centering
    \includegraphics[draft=false]{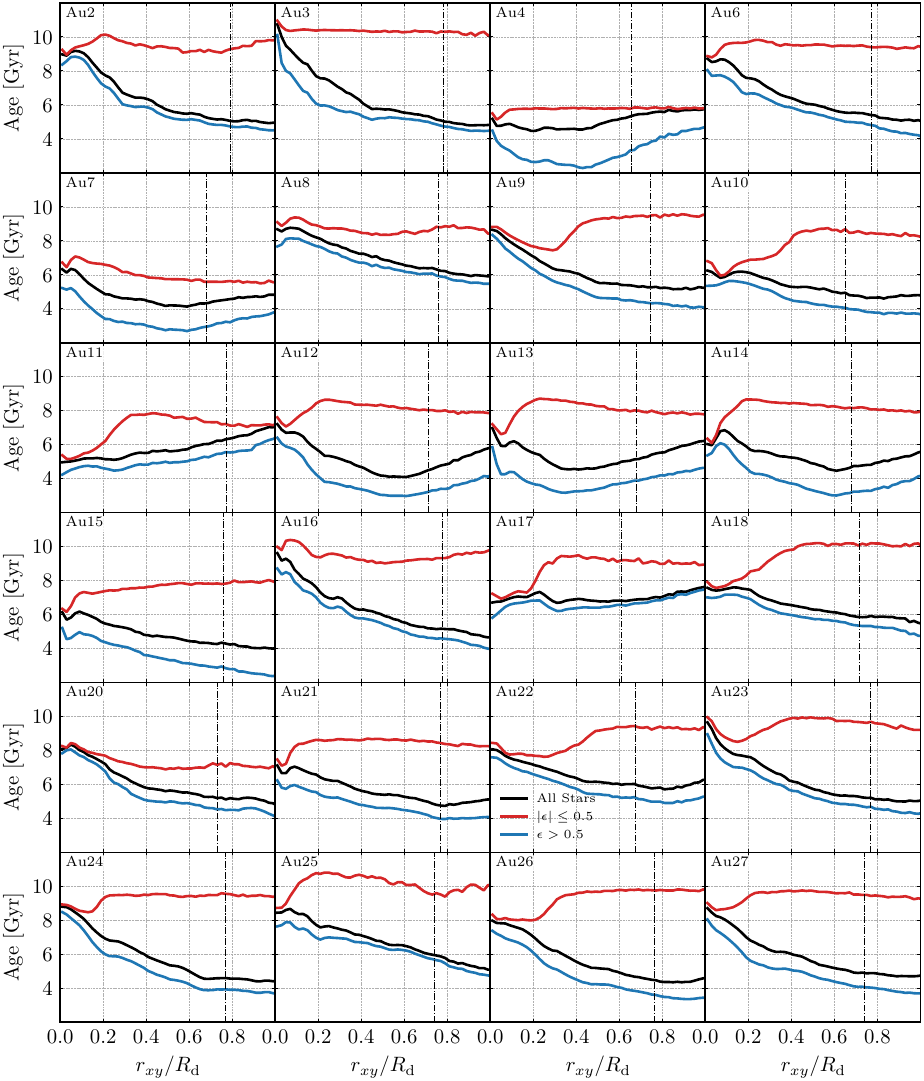}
    \caption{
        Average stellar age in Gyr as a function of radius on the disc plane (normalised to the disc radius) at $z=0$.
        Each panel shows the profile of a different galaxy and colours indicate the average age over all stars in black, stars with circularity below 0.5 in red (representative of the spheroid) and stars with circularities larger than 0.5 (representative of the disc) in blue, respectively.
        The vertical dashed-dotted line highlights the radius that encloses $80 \%$ of the stellar mass, following \protect\cite{Iza22}.
        Decreasing profiles indicate an ``inside-out'' distribution of stars, with older stars in the inner regions and younger stars in the outer regions.
    }
    \label{fig:binned_stellar_formation_time}
\end{figure*}

\subsubsection{Quantifying inside-out growth with the mass-weighted time}
\label{sec:mass-weighted-time}

After calculating the infall, outflow and net accretion rates, we would like to study the validity of the {\it inside-out} formation scenario within the context of cosmological simulations.
In this picture, usually accepted in the formation of MW-like galaxies, the stellar disc of galaxies forms their inner regions first and the outer regions at later times.
This is usually quantified by studying the radial profiles of stellar age with cosmic time, but as we showed in the previous paper of this series, there is a strong correlation between the amount of inflowing material and star formation rate since the latter is induced by the presence of new material that has fallen onto the disc.

We introduce a parameter that allows us to quantify the characteristic time at which a mass rate (the inflow rate, for example) is concentrated onto a specific disc region.
Then, we define the {\it mass-weighted time} $\tau$ as
\begin{equation}\label{eq:tau}
    \tau \equiv \frac{\int_{t_i}^{t_{0}} t \dot{M}(t) \mathrm{d}t}{\int_{t_i}^{t_{0}} \dot{M}(t) \mathrm{d}t},
\end{equation}
where $\dot{M}$ is a given mass rate (an amount of mass per unit time, used as a weight), $t$ is the cosmic time, $t_i$ is a reference time from which we start the calculation and $t_0$ is the age of the universe.
As shown in Paper I, stellar discs in the Auriga simulations are well defined only after the first $\sim 4\,\mathrm{Gyr}$ of evolution.
Hence, in what follows, we adopt $t_i=4\,\mathrm{Gyr}$ for the starting point in the integrals above.

Note that this parameter is essentially the mean value of the cosmic time weighted by the corresponding mass rate.
Therefore, for a given mass rate, $\tau$ indicates the mean time at which the mass rate is concentrated and, as we will show in Section~\ref{sec:results}, it can be used to quantify the inside-out behaviour of each galaxy.

\section{Results}
\label{sec:results}

In this Section we discuss whether the simulated discs in the Auriga galaxies formed from the inside-out.
In the context of galaxy formation, an inside-out scenario usually refers to an anti-correlation between the age of stars and their location onto the disc: in discs formed from the inside-out increasingly younger stars populate the outer regions of discs.
If stellar migration is not significant, inside-out will imply that stars in the inner (outer) regions of discs formed early (late).
As star formation is mainly determined by the availability of dense gas, a relation would then be expected between the amount of accreted gas as a function of radii and the star formation rate along the disc, which would be valid across cosmic history.
The following sections explore the inside-out formation scenario in the {\sc auriga} simulations, in terms of the stellar populations and the accretion history.

\subsection{Inside-out formation of the stellar discs}

Fig.~\ref{fig:binned_stellar_formation_time} shows the mean stellar age of stars as a function of projected radius at $z=0$ (normalised by the corresponding disc radii), for all stars (black lines) as well as separating stars into disc and spheroidal components (blue and red lines, respectively)\footnote{Disc stars were identified as those that are rotating in disc-like orbits. In particular, stars were considered as part of the disc if they have a circularity parameter $\epsilon>0.5$ and part of the spheroid if $\left| \epsilon \right| \leq 0.5$. As explained in Paper I (see also \citealt{Scannapieco2009}), the circularity parameter is defined as the ratio between the angular momentum of each star in the $z$-direction (i.e. perpendicular to the disc plane) and the angular momentum expected for a circular orbit at the star's radius. For this reason, disc stars will have $\epsilon\approx 1$ while a spheroid would be visible in the circularity distribution around $\epsilon\approx0$ if it is non-rotating.}.
In approximately $70\%$ of the galaxies, we observe a clear inside-out pattern for the disc components: the mean stellar age anti-correlates with radius.
This means that the oldest stars tend to be located in the inner regions, while they get increasingly younger towards larger radii.
In many other cases, we observe a similar behaviour up to about half the disc radius and the trend inverts in the outermost disc regions.
As expected, the spheroidal components in most cases show no correlation between mean stellar age and radius, and their populations are significantly older compared to the discs \citep{Monachesi2019}.
Nevertheless, some galaxies (e.g. Au9, Au11, Au17, and Au22)  show that stellar populations in the inner regions are -- on average -- younger compared to the outer regions.
This results from ongoing star formation in the inner parts of these galaxies\footnote{We discarded contamination of disc particles into the spheroidal component as a possible reason for this behaviour, which could in principle result from our simple separation of the disc and spheroidal components.}.

In the next sections, we investigate whether the behaviour observed for galaxies is consistent or not with an inside-out behaviour, and if this dichotomy could be explained by differences in their accretion patterns.

\subsection{Inflow and outflow rates}
\label{sec:results-inflowandoutflowrates}

\begin{figure*}
    \centering
    \includegraphics[draft=false]{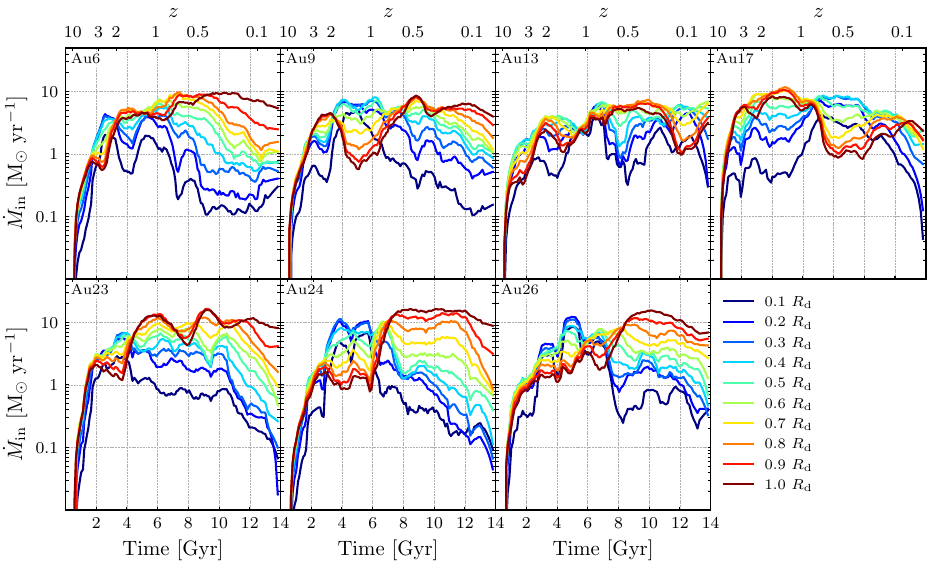}
    \caption{
        Temporal evolution of the gas inflow rate for different radial bins.
        Each line in this figure represents a different radial ring in the disc plane (the width is $0.1 R_\mathrm{d}$ for each ring); the first line (labelled as $0.1 R_\mathrm{d}$ in the legend), for example, represents the ring that extends from $r_{xy}=0$ to $r_{xy} = 0.1 R_\mathrm{d}$.
        The ten lines shown in each panel cover the complete radial extension of the disc for each galaxy.
        In this figure, the inside-out behaviour is evidenced by the relative heights of the curves for each radial bin; for Au26, for example, inflow rates are roughly similar for all bins before $\sim 7~\mathrm{Gyr}$ but higher rates are present for the outer regions near the present.
    }
    \label{fig:inflow_tracers}
\end{figure*}

The radial dependency of gas inflow and outflows rates onto the stellar discs of the 7 simulations including tracer particles are analysed in this Section.
As shown in Paper I, the inflow and outflow rates follow very similar patterns, with the former being always dominant over the latter.
For this reason, we only show the results obtained for the case of the inflow rates.
Fig.~\ref{fig:inflow_tracers} shows the evolution of the inflow rate, separated into various curves which represent the accretion onto different radial bins normalised with the corresponding disc radius ($R_\mathrm{d}$) of each galaxy.
Each radial ring is referenced by its maximum radius, and colour-coded using the colour map indicated in the legend.

In all cases except for Au13 and Au17, we observe that, for late times, the inflow rates are higher for larger radii, while the opposite occurs at early times (note that we focus on times $t\gtrsim 2~\mathrm{Gyr}$, i.e. when galaxies start to form their discs).
This inversion typically takes place in the range $t\sim 4-8\,\mathrm{Gyr}$.
At times $\gtrsim 10~\mathrm{Gyr}$, the infall rates for the different radial bins differ in more than an order of magnitude, while at earlier times the differences are less important.
Au13 and Au17, on the other hand, show a more complex behaviour.
In the first case, infalling rates tend to be higher in the outer regions except for times between $\sim11\,\mathrm{Gyr}$ and the current epoch.
In Au17 a similar inversion seems to take place around $\sim 6 ~\mathrm{Gyr}$ and up to $\sim 11 ~\mathrm{Gyr}$.
We come back to possible reasons for the behaviour of these galaxies in the next sections.

\begin{figure*}
    \centering
    \includegraphics[draft=false]{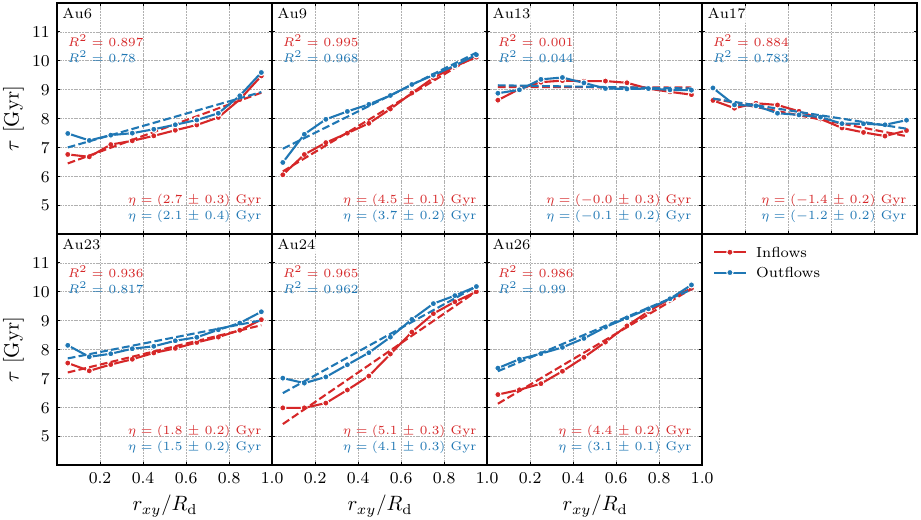}
    \caption{
        Mass-weighted time $\tau$ as a function of normalised radius for the re-simulated haloes.
        Each panel shows $\tau$ calculated adopting the inflow rate (red) and the outflow rate (blue) as a weight, as well as a linear fit for each case (dashed lines, corresponding colours).
        On the bottom right of each panel we indicate the slopes of the fits, along with its uncertainty (corresponding colours).
        Increasing (decreasing) profiles indicate that outer regions concentrate accretion at later (earlier) times than inner regions.
        Therefore, inside-out galaxies are evidenced by positive slopes.
    }
    \label{fig:mass_weighted_time_reruns}
\end{figure*}

The mass-weighted time $\tau$ defined in Section~\ref{sec:mass-weighted-time} can be used to estimate the typical times at which inflows occur for different radial bins and identify galaxies that are consistent with an inside-out formation scenario.
In Fig.~\ref{fig:mass_weighted_time_reruns}, we show $\tau$ as a function of ring radius for the 7 simulations previously discussed.
In this case, we include the results for the inflow rates of Fig.~\ref{fig:inflow_tracers}, as well as for the outflows.
All galaxies except for Au13 and Au17 exhibit a positive correlation between $\tau$ and radius for both inflows and outflows, confirming the results of the previous figure which suggested an inside-out pattern.
This means that, as we go from the inner to the outer regions of the discs, the dominant times at which inflows/outflows occur move from earlier ($\sim 5-8~\mathrm{Gyr}$) to later times ($\gtrsim 9~\mathrm{Gyr}$).
Au13 and Au17, on the other hand, show nearly constant and linearly decaying profiles, respectively.
For Au13, the accretion onto the disc seems to concentrate at $\sim 9~\mathrm{Gyr}$ regardless of the radius; in the case of Au17, accretion onto the inner (outer) region is concentrated at $\sim 8$--$9~\mathrm{Gyr}$ ($\sim 7$--$8~\mathrm{Gyr}$) (see Fig.~\ref{fig:mass_weighted_tau_net}).

Another useful quantity to measure the level of the inside-out pattern is the slope of the $\tau$-radius relation (the radius is normalised to the disc radius of each galaxy at any given time), which we refer to as the {\it inside-out parameter} $\eta$.
This parameter is calculated using a linear fit of the mass-weighted time as a function of radius and is given by
\begin{equation}
    \eta = \frac{\sum_{i=1}^N \left( r_i - \left< r \right> \right) \left( \tau_i - \left< \tau \right> \right)}{\sum_{i=1}^N \left( r_i - \left< r \right> \right)^2},
\end{equation}
where $r$ is the radius on the disc plane (normalised by the disc radius, as we consider in the following sections), $\tau$ is the mass-weighted time as defined in Eq.~\eqref{eq:tau}, $N$ is the amount of data points, and $\left< \right>$ denotes the average.
The values of $\eta$ are included in each panel of Fig.~\ref{fig:mass_weighted_time_reruns} along with their uncertainties.
The typical $\eta$ values are between $\sim2$ and $5\,$Gyr, and the $\eta$ values calculated from infalling and outflowing material are consistent with each other\footnote{We find similar $\eta$ values if we include radial flows in the calculation of the inflow/outflow rates. More important, the classification among inside-out versus no inside-out disc growth stays unchanged.}.
In the case of Au13 and Au17, whose behaviour differ from that of the rest of the galaxies, we find an $\eta\lesssim 0$ value.
In particular, $\eta\approx 0$ for Au13, both in terms of the inflow and outflow rates, while Au17 exhibits negative $\eta$ values.

It is worth noting that 5 out of 7 galaxies are consistent with an inside-out formation scenario which, as discussed above, provides a simple picture for the disc assembly of MW-type galaxies which is usually assumed in models of the formation of the MW.
Furthermore, galaxies identified as inside-out in terms of the stellar disc (see Fig.~\ref{fig:binned_stellar_formation_time}) are also identified as inside-out from the infall rates, and the same is valid for galaxies that do not seem to have formed in an inside-out manner.

Our simulations, which involve MW-mass galaxies with different formation histories, show that galaxy-to-galaxy variations can translate into different values of the inside-out parameter $\eta$ and that galaxies inconsistent with an inside-out behaviour can, in fact, also form \citep[see][]{Nuza2019}.
Estimating the relative fraction of galaxies with an inside-out pattern would require higher statistics.
In the next section we will assess the previous point using the net rates of the Auriga sample, for which we have a higher number of simulated galaxies.

\subsection{Net rates}
\label{sec:net_rates_results}

\begin{figure*}
    \centering
    \includegraphics[draft=false]{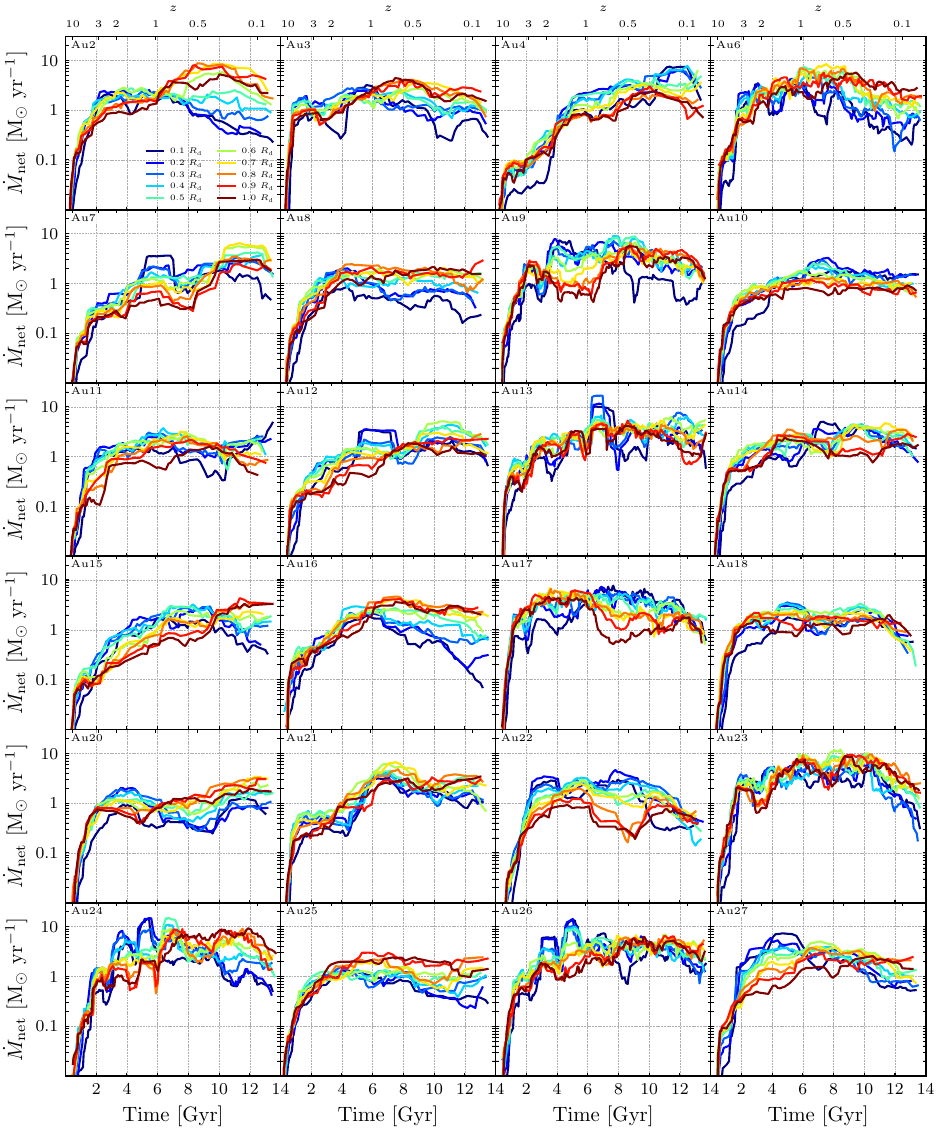}
    \caption{
        Temporal evolution of the gas (inflow-dominated) net rate (using cells) for different radial bins.
        Each line in this figure represents a different radial ring in the disc plane (the width is $0.1 R_\mathrm{d}$ for each ring); the first line (labelled as $0.1 R_\mathrm{d}$ in the legend), for example, represents the ring that extends from $r_{xy}=0$ to $r_{xy} = 0.1 R_\mathrm{d}$.
        The ten lines shown in each panel cover the complete radial extension of the disc for each galaxy.
    }
    \label{fig:net_cells}
\end{figure*}

\begin{figure*}
    \centering
    \includegraphics[draft=false]{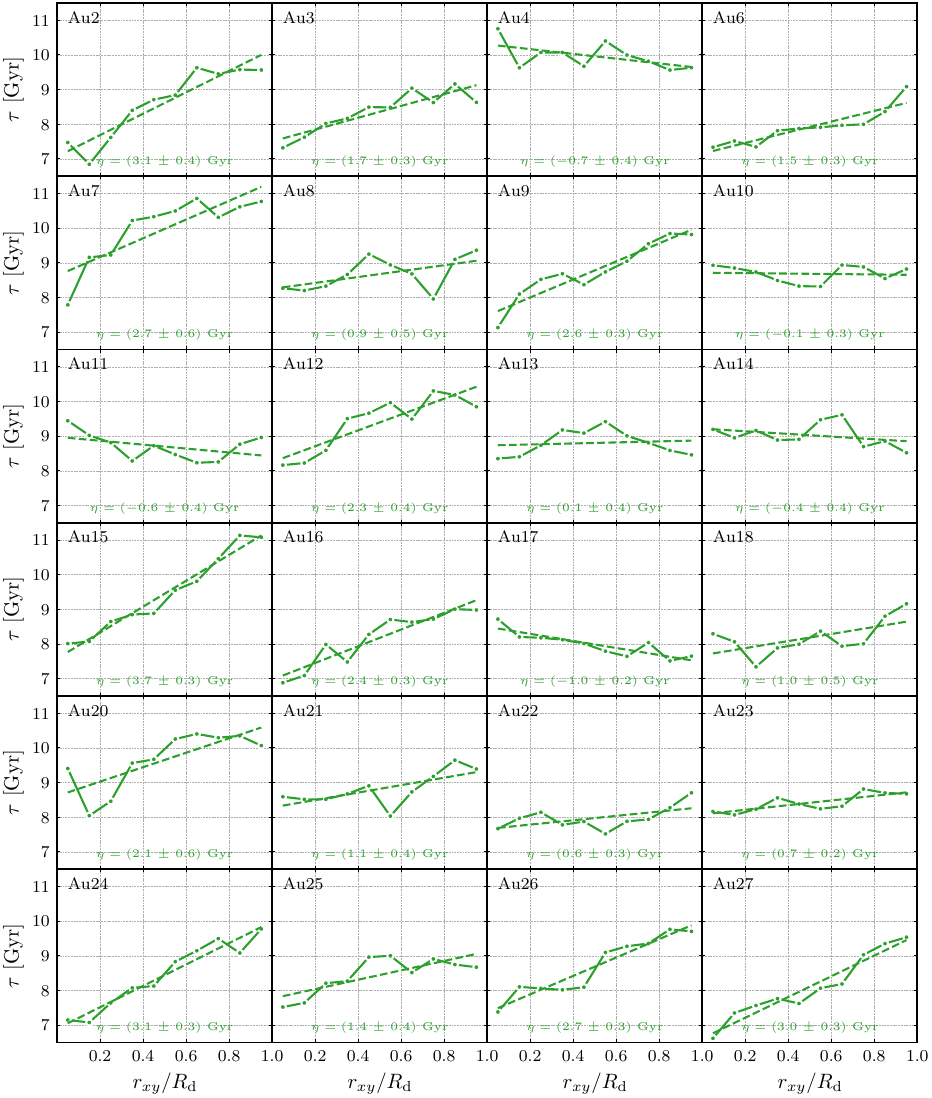}
    \caption{
        Mass-weighted time $\tau$ for all galaxies calculated adopting the net accretion rate as a weight, and a linear fit of the data (indicated as a dashed line).
        The slope of the fit, along with its uncertainty, is shown in each panel and can also be consulted in Table~\ref{tab:eta}.
    }
    \label{fig:mass_weighted_tau_net}
\end{figure*}

As explained in Paper I and Section~\ref{sec:inflow_outflow_rates_calculation}, the full Auriga sample can be used to estimate net inflow rates as a function of time\footnote{In \cite{Iza22} we have shown that the net inflow rates calculated from the cells are consistent with those estimated using the full information provided by the tracer particles (see Appendix~B of Paper I).}.
In this section, we investigate the net inflow rates as a function of the disc radius.
Fig.~\ref{fig:net_cells} shows the net accretion rates as a function of time for the simulated galaxies, separated in various radial bins covering the disc extension at each time.
Similarly to our results for the infall rates, we find that most of the galaxies are consistent with an inside-out pattern of the net accretion rates: accretion at early (late) times occurs preferentially at the innermost (outermost) radial bins.
This behaviour can be quantified using the mass-weighted time $\tau$ versus radius relation, which we show in Fig.~\ref{fig:mass_weighted_tau_net}, and the corresponding values for the slope of this relation, i.e., the already mentioned inside-out parameter $\eta$.

We find that the net accretion rates are in general compatible with an inside-out formation scenario, with $\eta$ values between $\approx 1$--$3~\mathrm{Gyr}$.
Some galaxies have, however, inside-out parameters consistent with zero or negative values.
In Table~\ref{tab:eta} we show $\eta$, its standard deviation $\sigma$ and the correlation parameter between the variables for all galaxies.
Using the $\eta$ parameter calculated with the net accretion rates, we classify galaxies as ``inside-out'' if $\eta\ge \eta_\mathrm{th}$, assuming $\eta_\mathrm{th}=0.8\,$Gyr as a threshold value.
With this choice, 16 out of the 24 galaxies fall in this category, wile the remaining 8 galaxies exhibit $\eta<\eta_\mathrm{th}$, 3 of which have $\eta<-0.5\,$Gyr\footnote{While the $\eta_\mathrm{th}$ value is somewhat arbitrary, the selected choice is the one that best represents the different behaviours observed in our sample. Furthermore, none of our conclusions are affected by this choice.}.

\begin{table*}
    \centering
    \caption{
        Inside-out parameter $\eta$ for each galaxy.
        We indicate the value of $\eta$, along with its uncertainty, for the inflow-dominated times of the net accretion rate (an its correlation coefficient), for the SFR, and for the inflows and outflows for the simulations that include tracer particles.
    }
    \label{tab:eta}
    \begin{tabular}{lccccccc}
        \hline
        Galaxy 	& $\eta_\mathrm{Net}$  & $R_\mathrm{Net}^2$ & $\eta_\mathrm{SFR}$ & $\eta_\mathrm{Inflows}$ & $\eta_\mathrm{Outflows}$ & Group \\
                    & [Gyr]                     & & [Gyr] & [Gyr] & [Gyr] & \cite{Iza22}      \\
        \hline
            Au2  & $3.1 \pm 0.4$  & 0.86 & $2.8 \pm 0.4$  & - & - & G1 \\
            Au3  & $1.7 \pm 0.3$  & 0.79 & $2.0 \pm 0.3$  & - & - & G1 \\
            Au4  & $-0.7 \pm 0.4$ & 0.29 & $-1.4 \pm 0.2$ & - & - & G2 \\
            Au6  & $1.5 \pm 0.3$  & 0.81 & $2.4 \pm 0.2$  & $2.7 \pm 0.3$ & $2.1 \pm 0.4$ & G1 \\
            Au7  & $2.7 \pm 0.6$  & 0.72 & $2.0 \pm 0.3$  & - & - & G2 \\
            Au8  & $0.9 \pm 0.5$  & 0.29 & $-0.3 \pm 0.2$ & - & - & G1 \\
            Au9  & $2.6 \pm 0.3$  & 0.89 & $4.9 \pm 0.4$  & $4.5 \pm 0.1$ & $3.7 \pm 0.2$ & G1 \\
            Au10 & $-0.1 \pm 0.3$ & 0.01 & $0.8 \pm 0.6$  & - & - & G1 \\
            Au11 & $-0.6 \pm 0.4$ & 0.19 & $-0.3 \pm 0.6$ & - & - & G1 \\
            Au12 & $2.3 \pm 0.4$  & 0.77 & $1.4 \pm 0.5$  & - & - & G2 \\
            Au13 & $0.1 \pm 0.4$  & 0.02 & $-1.4 \pm 0.4$ & $0.0 \pm 0.3$ & $-0.1 \pm 0.2$ & G1 \\
            Au14 & $-0.4 \pm 0.4$ & 0.12 & $-0.2 \pm 0.4$ & - & - & G1 \\
            Au15 & $3.7 \pm 0.3$  & 0.96 & $3.3 \pm 0.1$  & - & - & G2 \\
            Au16 & $2.4 \pm 0.3$  & 0.87 & $2.5 \pm 0.2$  & - & - & G1 \\
            Au17 & $-1.0 \pm 0.2$ & 0.75 & $-3.6 \pm 0.2$ & $-1.4 \pm 0.2$ & $-1.2 \pm 0.2$ & G1 \\
            Au18 & $1.0 \pm 0.5$  & 0.37 & $1.2 \pm 0.4$  & - & - & G1 \\
            Au20 & $2.1 \pm 0.6$  & 0.59 & $2.2 \pm 0.8$  & - & - & G2 \\
            Au21 & $1.1 \pm 0.4$  & 0.47 & $1.0 \pm 0.1$  & - & - & G1 \\
            Au22 & $0.6 \pm 0.3$  & 0.32 & $1.9 \pm 0.4$  & - & - & G1 \\
            Au23 & $0.7 \pm 0.2$  & 0.63 & $1.3 \pm 0.1$  & $1.8 \pm 0.2$ & $1.5 \pm 0.2$ & G1 \\
            Au24 & $3.1 \pm 0.3$  & 0.94 & $4.7 \pm 0.3$  & $5.1 \pm 0.3$ & $4.1 \pm 0.3$ & G1 \\
            Au25 & $1.4 \pm 0.4$  & 0.60 & $0.7 \pm 0.2$  & - & - & G1 \\
            Au26 & $2.7 \pm 0.3$  & 0.91 & $4.2 \pm 0.3$  & $4.4 \pm 0.2$ & $3.1 \pm 0.1$ & G1 \\
            Au27 & $3.0 \pm 0.3$  & 0.93 & $3.7 \pm 0.3$  & - & - & G1 \\
        \hline
    \end{tabular}
\end{table*}

As discussed in Sec.~\ref{sec:auriga}, it is worth noting that, in Paper I, we separated our galaxy sample into two groups: G1 formed by galaxies identified as MW analogues (systems with well-defined, stable discs at $z=0$ and no strong perturbations during the last $8~\mathrm{Gyr}$) and G2 composed of galaxies with well-defined discs at $z=0$, but with episodes of partial/total destruction of the discs after the first $4~\mathrm{Gyr}$ of evolution.
Given the small number statistics in group G2, we cannot confirm any systematic difference in the behaviour of galaxies between both groups (see Table~\ref{tab:eta}) in terms of their net accretion rates as a function of radius.
However, it is worth noting that $4$ out of the $5$ galaxies in G2 have very clear inside-out patterns, with $\eta>2~\mathrm{Gyr}$, and are among the galaxies with highest $\eta$ values.
However, we have checked that if we restrict the sample to times after the formation of the disc that survives up to the present time, the obtained $\eta$ values decrease and become similar to the values obtained for galaxies in G1.

\begin{figure*}
    \centering
    \includegraphics[draft=false]{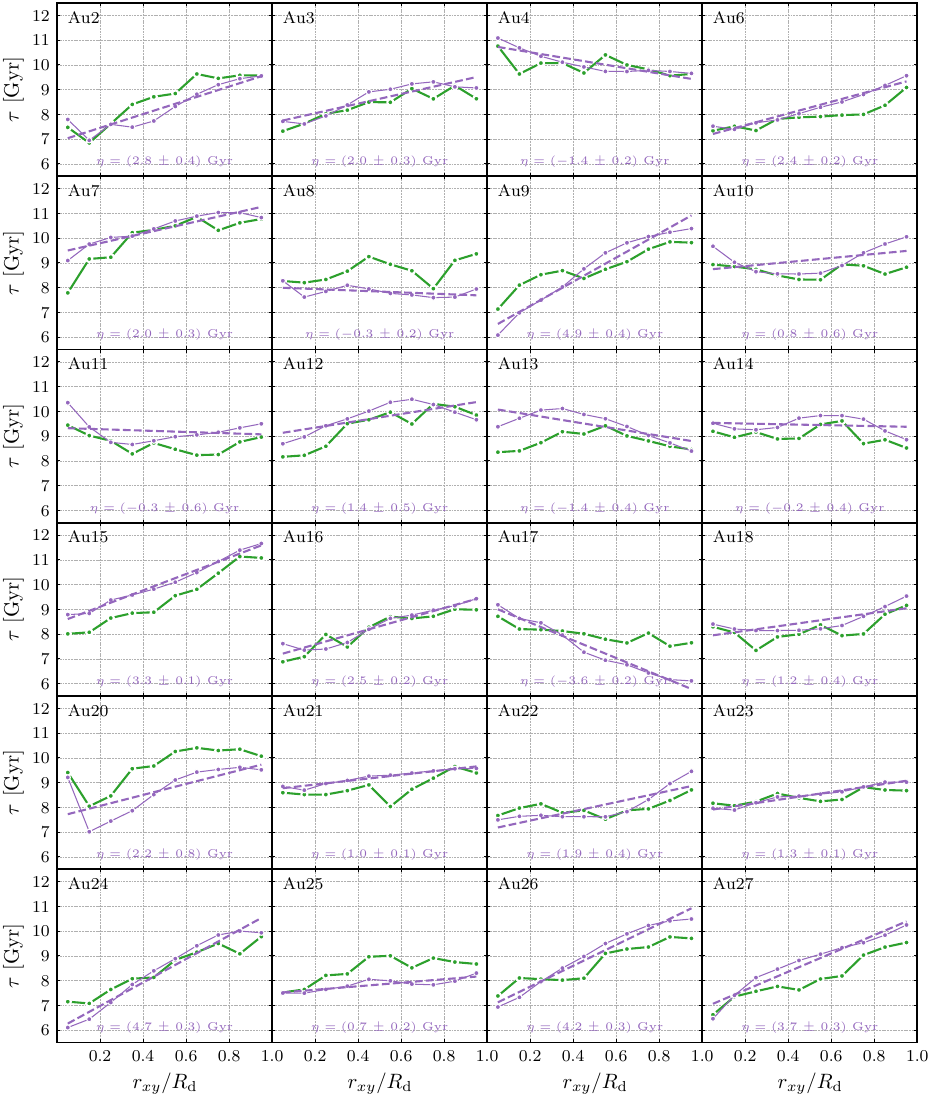}
    \caption{
        Mass-weighted time $\tau$ computed adopting the net accretion rate (green, repeated here for comparison) and the star formation rate (purple) as a weight.
        The purple dashed line shows a linear fit for the SFR curve and we indicate the slope and its uncertainty in each panel.
    }
    \label{fig:mass_weighted_time_net_sfr}
\end{figure*}

Among the galaxies that are incompatible with an inside-out formation scenario (Au4, Au10, Au11, Au13, Au14, Au17, Au22 and Au23), the most extreme cases in terms of the $\eta$ parameter are Au4 and Au17.
The latter has the highest slope (in absolute value), with $\eta=-1.0~\mathrm{Gyr}$, and has several distinctive features compared to the rest of the sample: it hosts a massive black hole, it has the highest accretion rate at early times, its stellar component formed very early on, is one of the most compact galaxies, and its environment is the densest one at very early times.
Au4, on the other hand, has an $\eta=-0.7~\mathrm{Gyr}$, but in this case it suffered a merger approximately $3~\mathrm{Gyr}$ ago,  shows a perturbed stellar distribution until the present time and is a very compact galaxy.
Au10, Au11, Au13 and Au14 have all suffered merger or interaction events and also show perturbed stellar distributions at late times, particularly after $10~\mathrm{Gyr}$.
In the case of Au11, a merger is occurring at $z=0$, as can be seen from Fig.~\ref{fig:density_maps_with_gas_1}.
It is worth noting that mergers might affect the orientation of the galaxy and the $\eta$ calculation.
Finally, Au22 and Au23 have $\eta$ values close to the assumed threshold of $0.8$, show quiescent formation histories and are similar to our sample of inside-out galaxies.

While most of the galaxies inconsistent with an inside-out behaviour have experienced recent mergers or have nearby satellites at the present day, no other systematics have been found in terms of virial masses, stellar masses, disc-to-total mass ratio, black hole mass, or even the presence/absence of a bar, as we show in Appendix~\ref{app:correlations}.
However, it is worth noting that the vast majority of them have high mean stellar formation times, that are systematically higher than those of the rest of the sample (Fig.~\ref{fig:eta_vs_props}).
This might suggest that discs have not yet fully formed or had enough time to grow and generate an inside-out pattern.
Furthermore, in  Appendix~\ref{app:dtt_inside_out} we show that, although these galaxies are not formally considered as ``inside-out'' globally (i.e., for our standard interval between $4~\mathrm{Gyr}$ and the present time), there are certain periods in which they show an inside-out behaviour, and that these periods often coincide with epochs of smooth disc growth, as opposed to intervals with an active merger activity.

\begin{figure*}
    \centering
    \includegraphics[draft=false]{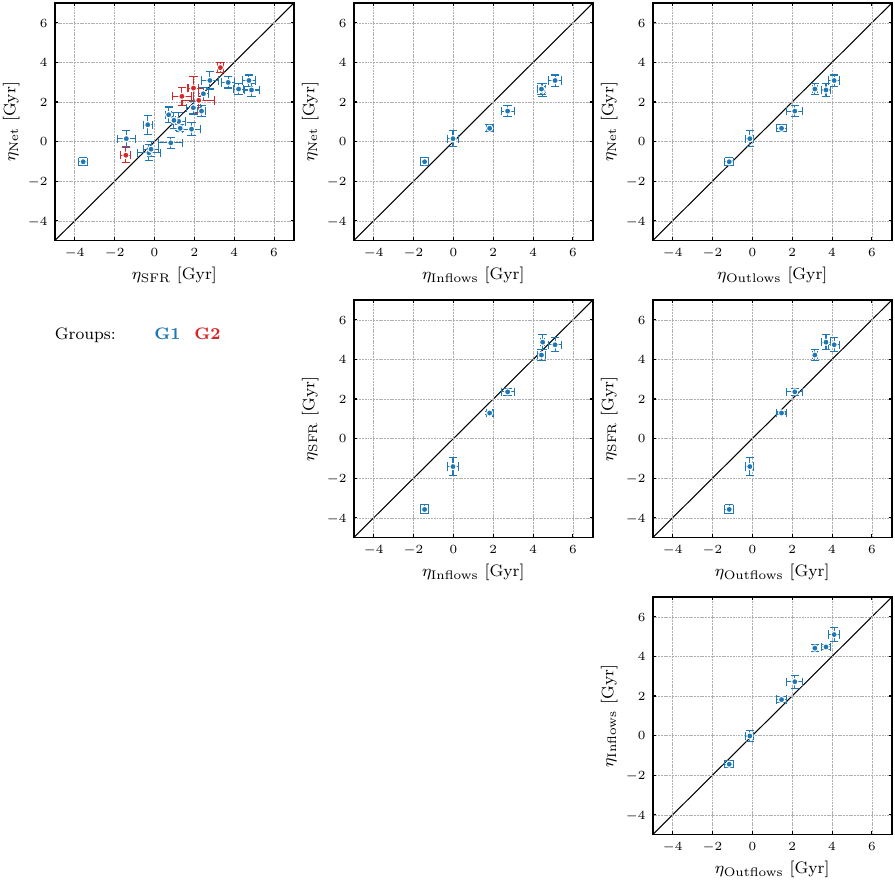}
    \caption{
        Correlations between the inside-out parameter $\eta$ calculated with different mass rates (net accretion rate, inflow rate, outflow rate and SFR).
        Each panel presents a different correlation, indicating the 1:1 line in black.
        Each point represents a given galaxy, colour-coded according to the group, as shown in the legend.
        The correlations shown here are close to the identity relation, with most deviations observed for low $\eta$ values.
    }
    \label{fig:correlations_in_out_net_sfr}
\end{figure*}

\subsection{Star formation rates and relation to accretion rates}

In this section, we investigate in more detail the relation between the star formation and the accretion rates along the discs.
Note that while we have already seen in Fig.~\ref{fig:binned_stellar_formation_time} which galaxies are compatible with an inside-out formation of the stellar disc, for consistency we also quantify this behaviour using the same methods as for the accretion rates.
Fig.~\ref{fig:mass_weighted_time_net_sfr} shows the radial profile of the mass-weighted time calculated adopting  the star formation rate as a weight of the simulated galaxies, along with the corresponding $\eta$ values.
For reference, we overplot the results corresponding to the $\tau$-radius relation of the net accretion rates.
We find that the star formation and net accretion rates are correlated, with very similar patterns and consistent $\eta$ values.
This result shows that gas inflows contribute with material which directly participates in the formation of new stars in the discs.
It is worth noting that, while in Paper I we also found a strong correlation between the evolution of the net accretion and star formation rates integrated over the disc extent, we find that these two quantities are also correlated in terms of their radial distributions along the discs.

In fact, we find a correlation not only among the $\eta$ parameters of the net accretion and star formation rates, $\eta_\mathrm{Net}$ and $\eta_\mathrm{SFR}$, respectively, but also with those of the infall and outflow rates, $\eta_\mathrm{Inflows}$ and $\eta_\mathrm{Outflows}$.
This is expected, as the combination of infall and outflows induced by feedback linked to star formation determines the net amount of gas which is deposited in the discs and can serve as a fuel for new stellar generations.
The correlations between the $\eta$ values obtained for the various rates is shown in Fig.~\ref{fig:correlations_in_out_net_sfr}, where we also indicate the identity relation.
Note that the infall and outflow rates can only be calculated for the simulations using tracer particles, and therefore the number of data points is smaller than for the full sample.
From this figure we can observe that all correlations are close to the identity relation, most notably the one determined by the net accretion rate and the SFR.
The most important deviations from the identity relation is found for galaxies with the lowest $\eta$ values.
We find no systematic differences in the behaviour of discs formed in our groups G1 and G2.

The results of this section suggest that the inside-out behaviour of gas accretion, at least for galaxies at the MW-mass scale, is responsible for the inside-out formation of the stellar discs.

\begin{figure*}
    \centering
    \includegraphics[draft=false]{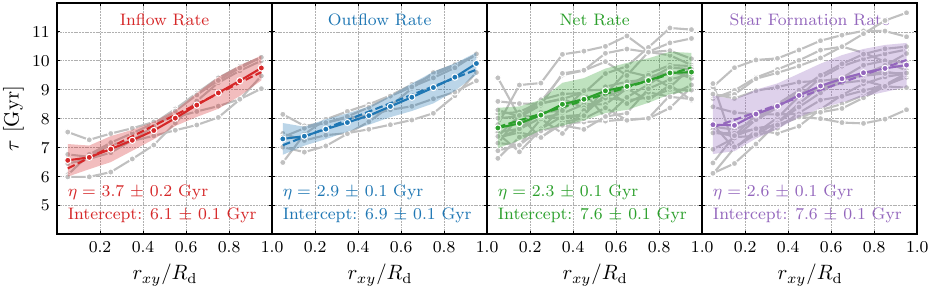}
    \caption{
        \emph{Left:} Mean mass-weighted time (red), calculated adopting the inflow rates as a weight, for all reruns excluding Au13 and Au17.
        The shaded region indicates the $\pm 1 \sigma$ region (this is the standard deviation of the data, not of the mean).
        The grey curves in the background correspond to each individual galaxy considered in the mean.
        \emph{Middle left:} Same as the panel on the left but for the outflow rates.
        \emph{Middle right:} Same as the panel on the left but for the net accretion rates calculated with the cells.
        In this case, we consider the fifteen galaxies that show an inside-out formation scheme: Au2, Au3, Au6, Au7, Au9, Au12, Au15, Au16, Au18, Au20, Au21, Au24, Au25, Au26, Au27.
        \emph{Right:} Same as the panel on the left but for the star formation rates.
        We include the same galaxies shown in the net accretion rate panel.
    }
    \label{fig:mass_weighted_time_mean}
\end{figure*}

In the remainder of this section and the rest of this work, we focus on the galaxies that are compatible with an inside-out behaviour, which represent roughly $70\%$ of our galaxy sample, and study their average behaviour in terms of the infall, outflow, net accretion and star formation rates along the discs.
Fig.~\ref{fig:mass_weighted_time_mean} shows the average mass-weighted time $\tau$ as a function of radius for the inflow, outflow, and net accretion rates and for the SFR.
The values obtained for the average inside-out parameters $\eta$, in all three cases, are also shown, together with the standard deviation.
The gas inflow rate presents the highest $\eta$ value, followed by the outflow and the net rates.
Note that the number of galaxies for which inflow and outflow rates can be calculated is significantly lower compared to the full sample.

Similarly to our previous analysis, the average mass-weighted time $\tau$ for all rates can be modelled as a function of the normalised disc radius by a simple linear relation of the form
\begin{equation}
    \tau (r_{xy}^*) = \eta r_{xy}^* + \tau_0,
\end{equation}
where $r_{xy}^*$ is the radius on the disc plane normalised by the size of the disc, $\eta$ is the slope (the inside-out parameter), and $\tau_0$ the intercept.
As long as $\eta$ is positive (or higher than a given threshold, as we previously commented) the behaviour can be thought of as ``inside-out''.
Table~\ref{tab:average_tau_results} shows the values of the parameters that result from a least squares linear fit of the mass-weighted time as a function of normalised radius, averaged over the G1 galaxies.
Note that the values presented in the table are not timescales related to an exponential decay but rather mean times around which the accretion is concentrated at any given normalised disc radius.

\begin{table}
    \centering
    \caption{
        Parameters of the linear fit the mass-weighted time as a function of normalised radius for the average behaviour of G1 (MW-like) galaxies.
        We indicate the value of the slope $\eta$ and intercept $\tau_0$, along with its uncertainties, for each of the mass rates we analysed.
    }
    \label{tab:average_tau_results}
    \begin{tabular}{lcc}
        \hline
        Mass rate               & $\eta$ (Slope)        & $\tau_0$ (Intercept)  \\
                                & [Gyr]                 & [Gyr]                 \\
        \hline
            Inflow Rate         & $3.7 \pm 0.2$         & $6.1 \pm 0.1$         \\
            Outflow Rate        & $2.9 \pm 0.1$         & $6.9 \pm 0.1$         \\
            Net Rate            & $2.3 \pm 0.1$         & $7.6 \pm 0.1$         \\
            Star Formation Rate & $2.6 \pm 0.1$         & $7.6 \pm 0.1$         \\
        \hline
    \end{tabular}
\end{table}

\section{Comparison with chemical evolution models}
\label{sec:cem_comparison}

\begin{figure*}
    \centering
    \includegraphics[draft=false]{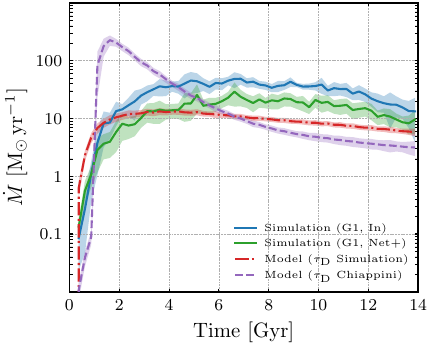}
    \includegraphics[draft=false]{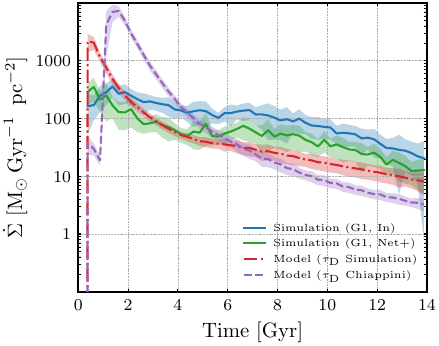}
    \includegraphics[draft=false]{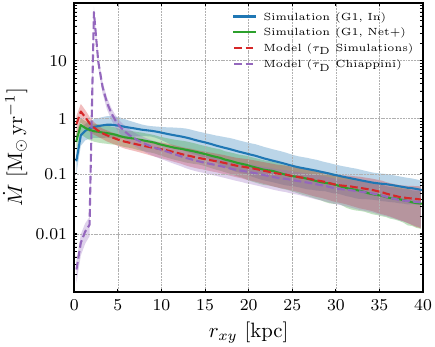}
    \includegraphics[draft=false]{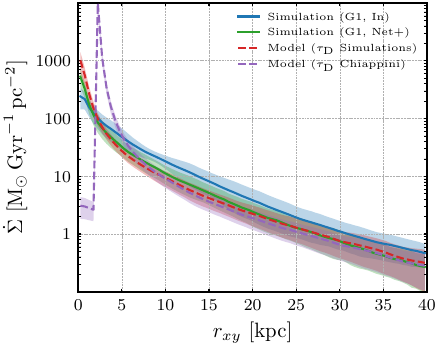}
    \caption{
        Comparison between the accretion calculated from the simulations and the Double Infall Model, commonly used in CEMs to model the accretion onto galactic discs.
        The curves are the results of the simulation (for the inflow-dominated times of the net accretion rate in green and for the inflow rate in blue) and the results of the model (using disc time-scale calculated from simulation data in red and one commonly used in CEMs in purple).
        We show the $\pm 3 \sigma$ regions for each line in the corresponding colour.
        The top left panel shows the disc-integrated rate as a function of time, the top right panel shows the disc-integrated flux as a function of time, the bottom left panel shows the time-integrated rate as a function of disc radius and the bottom right panel shows the time-integrated flux as a function of disc radius.
        The calculation of each of these values is detailed in the text.
    }
    \label{fig:cem_comparison}
\end{figure*}

Given the average behaviour\footnote{Note that, in general, the average behaviour of the mass-weighted time $\tau$ presented in Fig.~\ref{fig:mass_weighted_time_mean} does not show strong deviations from that of each individual galaxy; therefore, in this section, we assume average functions for a simpler comparison.} of gas accretion rates in our sample of simulated MW-like galaxies in G1 we would like to compare the results obtained from simulations with those of 
chemical evolution models (CEMs) which are semi-analytical schemes developed with the aim of providing a description for the distribution of metals in galaxies and, particularly, in the MW.
These models employ a series of analytical hypotheses that simulate certain aspects of the process of galaxy formation and evolution such as star formation and, of particular interest in this work, the accretion of gas onto the disc.
One of the scenarios that have been proposed in the literature is the so-called ``Double Infall Model'' (hereafter DIM), built on top of the general assumption that the gas accretion process in disc-like galaxies takes place in two distinct phases \citep[e.g.][]{Chiappini01}.

In this context, the accretion of gas is given by the sum of two components linked to the formation of the stellar halo, bulge, and thick disc, on one hand, and to the thin disc, on the other.
The rate of mass accretion per unit area as a function of cosmic time $t$ and disc radius $R$ is usually expressed as \citep[see e.g.][]{Matteucci12}:
\begin{equation}
    \dot{\Sigma}(R, t) = A(R) \mathrm{e}^{-t/ \tau_\mathrm{H}(R)} + B(R) \mathrm{e}^{-(t-t_\mathrm{max})/ \tau_\mathrm{D}(R)},
    \label{CEMaccretion}
\end{equation}
where the first term represents the accretion linked to the spheroidal/thick disc components and the second represents the accretion onto the thin disc.
The radial functions $A(R)$ and $B(R)$ are chosen to reproduce the total mass surface density of each stellar component at the present time (see below), $\tau_\mathrm{H}$ and $\tau_\mathrm{D}$ are typical timescales for the formation of the spheroidal and thin disc components which may depend on $R$, and $t_\mathrm{max}$ is the time at which the maximum accretion value occurs for the disc.

These models take into account some common assumptions that are motivated by observational results.
In particular, the formation timescale of the spheroidal component's half mass, $\tau_\mathrm{H}$, is assumed to be independent of radius and equal to $0.8~\mathrm{Gyr}$ since the formation of the bulge is believed to occur in a violent, rapid process at early times.
Similarly, the time of maximum accretion, $t_\mathrm{max}$, is commonly chosen to be either $1$ or $2~\mathrm{Gyr}$ and corresponds to the end of the spheroidal/thick disc phase.

The formation timescale of the thin disc, $\tau_\mathrm{D}$, is chosen in consistency with the inside-out formation scenario.
This is done by taking the characteristic rate of gas accretion as an increasing function of the disc radius in order to properly reproduce the metallicity distribution of stars in the solar neighbourhood and the bulge.
In \cite{Chiappini01} this dependency is obtained by assuming the following linear function:
\begin{equation}
    \tau_\mathrm{D}(R) = \left( 1.033 \frac{R}{\mathrm{kpc}} - 1.267 \right)\,\mathrm{Gyr}.
    \label{TauChiappini}
\end{equation}

As already mentioned, the normalisation profiles $A(R)$ and $B(R)$ are chosen to reproduce, respectively, the total mass surface density of the spheroidal and thin disc components at $z=0$ or present-day time $t_\mathrm{G}$.
It can be shown that the first normalisation function can be written as
\begin{equation}
    A(R) = \frac{\Sigma_\mathrm{H}(R, t_\mathrm{G})}{\tau_\mathrm{H} \left( 1- \mathrm{e}^{-t_\mathrm{G} / \tau_\mathrm{H}} \right)},
    \label{Ar}
\end{equation}
where $\Sigma_\mathrm{H}(R, t_\mathrm{G})$ is the mass surface density of the spheroidal component as a function of radius at $z=0$ assuming that $\tau_\mathrm{H} (R)$ is independent of $R$.
For the thin disc, the $B(R)$ function can be written as
\begin{equation}
    B(R) = \frac{\Sigma_\mathrm{D}(R, t_\mathrm{G})}{\tau_\mathrm{D}(R) \left( 1 - \mathrm{e}^{-(t_\mathrm{G} - t_\mathrm{max})/ \tau_\mathrm{D}(R)} \right)}.
    \label{Br}
\end{equation}
As before, $\Sigma_\mathrm{D}(R, t_\mathrm{G})$ is the mass surface density of the thin disc component as a function of radius at $z=0$.
To avoid nonphysical results when using Eq.~(\ref{TauChiappini}) for in this formula\footnote{Note that the timescale defined as $\tau_\mathrm{D}(R) = \left( 1.033\, R / \mathrm{kpc} - 1.267 \right)\,\mathrm{Gyr}$ yields negative values for $R < 1.227 ~\mathrm{kpc}$.}, we simply set the normalisation constant $B(R)$ to zero for $R \leq 2~\mathrm{kpc}$, thus eliminating the disc accretion component from the model in the very central regions.

To make a direct comparison between simulations and the DIM, we recomputed the inflow/net accretion rates onto the Auriga galaxies using a fixed number of bins in physical coordinates between $0$ and $40~\mathrm{kpc}$, thus keeping the bin width constant.
The mass surface density of the disc was computed using all mass of stars and gas inside a cylinder of $40~\mathrm{kpc}$ radius and disc height as defined in \cite{Iza22}.
For the spheroidal components, we considered all remaining mass in the subhalo that is not part of the disc.
In Eqs.~(\ref{CEMaccretion}) and~(\ref{Br}) we used both the \cite{Chiappini01} relation and those calculated directly from our simulations as described in Section~\ref{sec:results} for gas accretion timescales $\tau_\mathrm{D}(R)$.
Since the accretion rates of the Auriga galaxies do not always follow a clear exponential decay, obtaining a timescale for each galactic disc can be a challenging task, specially as a function of disc radius.
For this reason, instead of computing the mass-weighted time as previously described, we calculated an accretion timescale for {\it each} radial ring in order to find the overall trend for $\tau_\mathrm{D}(R)$.
For the remaining parameters, we used the fiducial CEM values and checked that when varied within a reasonable range for the corresponding physical constraints there are no significant variations that change the overall conclusions of this comparison.

Fig.~\ref{fig:cem_comparison} shows the comparison between the accretion rates obtained from simulations and the results of the DIM.
In all cases, we show an average of the inflow-dominated times of the inflow/net accretion rates corresponding to galaxies in the group G1 defined in \cite{Iza22} (i.e. the MW-like galaxy sample) together with the $\pm 3 \sigma$ regions.
Three curves are shown in the plot: the mean accretion rates obtained from simulations (green), the CEM model using the $\tau_\mathrm{D}$ defined from simulation data (red), and the CEM model but using one of the fiducial $\tau_\mathrm{D}$ relations usually adopted in CEMs (purple).

The panels on the first row show the temporal evolution of the disc-integrated rates.
Here, in order to perform the radial integration, we consider galactic discs between $0$ and the disc radius $R_\mathrm{d}$ calculated in \cite{Iza22}.
On the left, we show the rates in units of solar masses per year, whereas on the right we divide the rate at each time by the corresponding disc areas and the evolution is presented in flux units of solar masses per gigayear per square parsec.
The panels on the second row, on the other hand, show the time-integrated rates as a function of the radius on the disc plane up to $40~\mathrm{kpc}$, both as mass rates (left) and mass flux (right).

As it can be seen in Fig.~\ref{fig:cem_comparison}, generally speaking, the net accretion computed from the simulation and the infall accretion resulting from the DIM show an order of magnitude agreement displaying a similar dependence with time and radius when radial- and time-integrated values are considered.
When focusing on the radially-integrated values, the differences observed in the evolution of the rates are less than one order of magnitude for times greater than $\sim4~\mathrm{Gyr}$, after which the discs of the Auriga galaxies can be considered as ``well-developed'' in terms of the disc-to-total mass ratio.
Furthermore, there is considerable agreement on the slope of the temporal decay of rates, which are similar at late times in both cases.
On the time-integrated panels, there is a considerably better agreement between the three curves shown.
This is especially evident after the first $\sim4~\mathrm{Gyr}$ of evolution, where all curves are well within the $3\sigma$ regions shown in each panel.

However, some major differences arise at early times and in the inner regions.
In the radially-integrated panels, there is a considerable increase in the rates calculated with the fiducial disc timescale of the DIM (purple line), reaching values of order $100 ~\mathrm{M}_\odot \, \mathrm{yr}^{-1}$.
A similar behaviour can be observed in the time-integrated panels, where the accretion rates reach values of order $10 ~ \mathrm{M}_\odot \, \mathrm{yr}^{-1}$ near $\sim 2 ~\mathrm{kpc}$ in the rate case.

Finally, we stress that the sharp discontinuity seen in the time-integrated panels of Fig.~\ref{fig:cem_comparison} at $2~\mathrm{kpc}$ and, owing to the size of the disc decreasing to almost zero at the beginning of the simulation, as a sharp decrease in the accretion of the model at times smaller than $\sim 2~\mathrm{Gyr}$.
We note, however, that at times less than $4~\mathrm{Gyr}$ the galactic discs are not necessarily formed (see \citealt{Iza22}) and, although we considered the inner regions of the galaxy as part of the disc for the calculation, they tend to be dominated by bulges instead of rotationally supported structures \citep{Grand2017}.

This comparison, finally, shows similar values between the accretion rates calculated from simulation and the results obtained from the DIM.
Time- and disc-integrated values show similar trends when analysing the radial and temporal dependency, respectively.
Most differences between the model and the simulations arise at times or regions when or where the disc may not be completely defined.

\section{Conclusions}
\label{sec:conclusions}

In this work we have investigated the inflow, outflow and net accretion rates onto the discs of MW-mass galaxies, simulated in a full cosmological setting in the context of the $\Lambda$CDM model.
The simulations are part of the {\sc Auriga} Project \citep{Grand2017}, and were run with the magnetohydrodynamical, moving-mesh code {\sc arepo} \citep{Springel2010}.
This is the second paper of a series where we analyse the gas flows at the disc-halo interface.
In Paper I, we investigated the temporal dependencies of the net accretion rates, integrated onto the radial extent of the discs, as well as for a subset of $9$ simulations which included a treatment of tracer particles that allowed us to separately calculate the inflow and outflow rates onto the discs.
Here, we focus on the radial dependencies of the inflow, outflow, net accretion and star formation rates onto the disc region, with the aim of verifying whether the usual assumption of inside-out formation of discs for MW-like galaxies is valid in our set of simulations.
For this, we excluded from our analysis galaxies whose discs are strongly perturbed during evolution, leaving a total of $24$ simulated galaxies, $7$ of which have also been run with tracer particles.

More than $70\%$ of the simulated galaxies were found to be consistent with an inside-out formation scenario of the stellar discs, showing an anti-correlation between the age of stars and their radial location along the disc.
In order to investigate the origin of this pattern, we calculated the gas flow and star formation rates as a function of radius, and characterised their radial dependencies using the $\tau$ parameter which measures, for various radial bins, the dominant times of each considered rate.
The slope of this relation, denoted as the ``inside-out parameter'' $\eta$, allows us to identify systems compatible with an inside-out behaviour and it was computed for the infall and outflow rates $(\eta_\mathrm{Inflows}$ and $\eta_\mathrm{Outflows}$, respectively), the net accretion rate ($\eta_\mathrm{Net}$) and the star formation rate ($\eta_\mathrm {SFR}$).

A very tight correlation was found between $\eta_\mathrm{Net}$ and $\eta_\mathrm{SFR}$, as well as between these two and $\eta_\mathrm{Inflows}$ and $\eta_\mathrm{Outflows}$ for the simulations with tracer particles.
This is expected, as the connections between the star formation and gas flow rates are a natural consequence of the interplay between star formation, the corresponding feedback, and accretion to the discs which determine, at each time and radius, the amount of gas available for star formation.
Assuming a conservative choice for the identification of galaxies consistent with an inside-out behaviour ($\eta\geq0.8\,$Gyr), we found that the discs of $16$ out of the $24$ galaxies of the sample follow this trend.
Our results show that the inside-out formation of discs is a natural consequence of an analogous accretion process in MW-mass galaxies formed in the context of the $\Lambda$CDM model.

For galaxies whose discs do not show an inside-out growth, we found that the relation between accretion/star formation rates and radius is in most cases flat, and, in a few cases, these have negative slopes pointing to an ``outside-in'' formation law.
Most of these galaxies show an active merger activity at late times suggesting that the associated discs did not have enough time to grow in an inside-out manner.
The number of galaxies in either of these cases is, however, too small in our galaxy sample to draw any strong conclusion concerning this at this point, but this is certainly worth exploring in future works.

The accretion laws obtained for the simulated galaxies are in relative good agreement with the assumptions usually made in CEMs, in particular in relation to the inside-out growth.
Comparisons between the gas accretion rates obtained in our simulations with those expected from classical CEMs for a set of galaxy parameters consistent with our MW-like galaxy sample (G1) show similar slopes for the accretion rate (radially-integrated along the discs) versus cosmic time but with simulated results higher by roughly a factor of 2.
At very early times, however, when the galactic discs are not yet formed, differences between CEMs and simulations are larger.
The accretion rates (temporally-integrated across galactic history) versus disc radius show more compatible results except in the very inner regions.
The relative agreement between the CEM model and the simulations is owing to the fact that the CEM is fed with the mean galactic parameters describing our MW-like galaxy sample, thus validating the general picture of the DIM.
However, when comparing simulations to the CEM expectation using the $\tau_{\rm D}(R)$ relation adopted in \cite{Chiappini01} the differences between the model and simulations are larger.
This indicates that, although there is always a positive correlation between the accretion timescale $\tau$ and disc radius, and that the rate of gas flows in the Solar neighbourhood is similar for inside-out galaxies in CEMs and simulations, the trends are quantitatively different.
Our analysis suggests that galaxy formation simulations of MW-like galaxies in a cosmological context predict a less steep rise in the accretion timescales as one moves farther from the galactic centre.

\section*{Acknowledgements}

SEN and CS are members of CONICET.
They acknowledge funding from Agencia Nacional de Promoci\'on Cient\'{\i}fica y Tecnol\'ogica (PICT 2021-GRF-TI-00290).
SEN acknowledges financial support from CONICET through the project PIBAA R73734.
FAG acknowledges financial support from FONDECYT Regular 1211370, and from the Max Planck Society through a Partner Group grant.
FAG gratefully acknowledges support by the ANID BASAL project FB210003.


\section*{Data availability}

The scripts and plots for this article will be shared on reasonable request to the corresponding author.
The \textsc{arepo} code is publicly available \citep{Weinberger2020}.



\bibliographystyle{mnras}
\bibliography{bibliography} 

\begin{thebibliography}{}
\makeatletter
\relax
\def\mn@urlcharsother{\let\do\@makeother \do\$\do\&\do\#\do\^\do\_\do\%\do\~}
\def\mn@doi{\begingroup\mn@urlcharsother \@ifnextchar [ {\mn@doi@}
  {\mn@doi@[]}}
\def\mn@doi@[#1]#2{\def\@tempa{#1}\ifx\@tempa\@empty \href
  {http://dx.doi.org/#2} {doi:#2}\else \href {http://dx.doi.org/#2} {#1}\fi
  \endgroup}
\def\mn@eprint#1#2{\mn@eprint@#1:#2::\@nil}
\def\mn@eprint@arXiv#1{\href {http://arxiv.org/abs/#1} {{\tt arXiv:#1}}}
\def\mn@eprint@dblp#1{\href {http://dblp.uni-trier.de/rec/bibtex/#1.xml}
  {dblp:#1}}
\def\mn@eprint@#1:#2:#3:#4\@nil{\def\@tempa {#1}\def\@tempb {#2}\def\@tempc
  {#3}\ifx \@tempc \@empty \let \@tempc \@tempb \let \@tempb \@tempa \fi \ifx
  \@tempb \@empty \def\@tempb {arXiv}\fi \@ifundefined
  {mn@eprint@\@tempb}{\@tempb:\@tempc}{\expandafter \expandafter \csname
  mn@eprint@\@tempb\endcsname \expandafter{\@tempc}}}

\bibitem[\protect\citeauthoryear{{Aumer}, {White}, {Naab}  \&
  {Scannapieco}}{{Aumer} et~al.}{2013}]{Aumer2013}
{Aumer} M.,  {White} S. D.~M.,  {Naab} T.,   {Scannapieco} C.,  2013, \mn@doi
  [\mnras] {10.1093/mnras/stt1230}, \href
  {https://ui.adsabs.harvard.edu/abs/2013MNRAS.434.3142A} {434, 3142}

\bibitem[\protect\citeauthoryear{{Aumer}, {White}  \& {Naab}}{{Aumer}
  et~al.}{2014}]{Aumer2014}
{Aumer} M.,  {White} S. D.~M.,   {Naab} T.,  2014, \mn@doi [\mnras]
  {10.1093/mnras/stu818}, \href
  {https://ui.adsabs.harvard.edu/abs/2014MNRAS.441.3679A} {441, 3679}

\bibitem[\protect\citeauthoryear{{Boissier} \& {Prantzos}}{{Boissier} \&
  {Prantzos}}{1999}]{Boissier99}
{Boissier} S.,  {Prantzos} N.,  1999, \mn@doi [\mnras]
  {10.1046/j.1365-8711.1999.02699.x}, \href
  {http://adsabs.harvard.edu/abs/1999MNRAS.307..857B} {307, 857}

\bibitem[\protect\citeauthoryear{{Chiappini}, {Matteucci}  \&
  {Romano}}{{Chiappini} et~al.}{2001}]{Chiappini01}
{Chiappini} C.,  {Matteucci} F.,   {Romano} D.,  2001, \mn@doi [\apj]
  {10.1086/321427}, \href
  {https://ui.adsabs.harvard.edu/abs/2001ApJ...554.1044C} {554, 1044}

\bibitem[\protect\citeauthoryear{{DeFelippis}, {Genel}, {Bryan}  \&
  {Fall}}{{DeFelippis} et~al.}{2017}]{DeFelippis2017}
{DeFelippis} D.,  {Genel} S.,  {Bryan} G.~L.,   {Fall} S.~M.,  2017, \mn@doi
  [\apj] {10.3847/1538-4357/aa6dfc}, \href
  {https://ui.adsabs.harvard.edu/abs/2017ApJ...841...16D} {841, 16}

\bibitem[\protect\citeauthoryear{{Fall} \& {Efstathiou}}{{Fall} \&
  {Efstathiou}}{1980}]{Fall1980}
{Fall} S.~M.,  {Efstathiou} G.,  1980, \mn@doi [\mnras]
  {10.1093/mnras/193.2.189}, \href
  {https://ui.adsabs.harvard.edu/abs/1980MNRAS.193..189F} {193, 189}

\bibitem[\protect\citeauthoryear{{Fragkoudi}, {Grand}, {Pakmor}, {Springel},
  {White}, {Marinacci}, {Gomez}  \& {Navarro}}{{Fragkoudi}
  et~al.}{2021}]{Fragkoudi2021}
{Fragkoudi} F.,  {Grand} R.~J.~J.,  {Pakmor} R.,  {Springel} V.,  {White}
  S.~D.~M.,  {Marinacci} F.,  {Gomez} F.~A.,   {Navarro} J.~F.,  2021, \mn@doi
  [\aap] {10.1051/0004-6361/202140320}, \href
  {https://ui.adsabs.harvard.edu/abs/2021A&A...650L..16F} {650, L16}

\bibitem[\protect\citeauthoryear{{Frankel}, {Sanders}, {Rix}, {Ting}  \&
  {Ness}}{{Frankel} et~al.}{2019}]{Frankel2019}
{Frankel} N.,  {Sanders} J.,  {Rix} H.-W.,  {Ting} Y.-S.,   {Ness} M.,  2019,
  \mn@doi [\apj] {10.3847/1538-4357/ab4254}, \href
  {https://ui.adsabs.harvard.edu/abs/2019ApJ...884...99F} {884, 99}

\bibitem[\protect\citeauthoryear{{Genel}, {Vogelsberger}, {Nelson}, {Sijacki},
  {Springel}  \& {Hernquist}}{{Genel} et~al.}{2013}]{Genel2013}
{Genel} S.,  {Vogelsberger} M.,  {Nelson} D.,  {Sijacki} D.,  {Springel} V.,
  {Hernquist} L.,  2013, \mn@doi [\mnras] {10.1093/mnras/stt1383}, \href
  {https://ui.adsabs.harvard.edu/abs/2013MNRAS.435.1426G} {435, 1426}

\bibitem[\protect\citeauthoryear{{G{\'o}mez} et~al.,}{{G{\'o}mez}
  et~al.}{2017}]{Gomez2017}
{G{\'o}mez} F.~A.,  et~al., 2017, \mn@doi [\mnras] {10.1093/mnras/stx2149},
  \href {https://ui.adsabs.harvard.edu/abs/2017MNRAS.472.3722G} {472, 3722}

\bibitem[\protect\citeauthoryear{{Grand} et~al.,}{{Grand}
  et~al.}{2017}]{Grand2017}
{Grand} R. J.~J.,  et~al., 2017, \mn@doi [\mnras] {10.1093/mnras/stx071}, \href
  {https://ui.adsabs.harvard.edu/abs/2017MNRAS.467..179G} {467, 179}

\bibitem[\protect\citeauthoryear{{Grand} et~al.,}{{Grand}
  et~al.}{2019}]{Grand2019}
{Grand} R. J.~J.,  et~al., 2019, \mn@doi [\mnras] {10.1093/mnras/stz2928},
  \href {https://ui.adsabs.harvard.edu/abs/2019MNRAS.490.4786G} {490, 4786}

\bibitem[\protect\citeauthoryear{{Guedes}, {Callegari}, {Madau}  \&
  {Mayer}}{{Guedes} et~al.}{2011}]{Guedes2011}
{Guedes} J.,  {Callegari} S.,  {Madau} P.,   {Mayer} L.,  2011, in Galaxy
  Formation. p.~P173

\bibitem[\protect\citeauthoryear{{Hopkins} et~al.,}{{Hopkins}
  et~al.}{2018}]{Hopkins2018}
{Hopkins} P.~F.,  et~al., 2018, \mn@doi [\mnras] {10.1093/mnras/sty1690}, \href
  {https://ui.adsabs.harvard.edu/abs/2018MNRAS.480..800H} {480, 800}

\bibitem[\protect\citeauthoryear{{Iza}, {Scannapieco}, {Nuza}, {Grand},
  {G{\'o}mez}, {Springel}, {Pakmor}  \& {Marinacci}}{{Iza}
  et~al.}{2022}]{Iza22}
{Iza} F.~G.,  {Scannapieco} C.,  {Nuza} S.~E.,  {Grand} R. J.~J.,  {G{\'o}mez}
  F.~A.,  {Springel} V.,  {Pakmor} R.,   {Marinacci} F.,  2022, \mn@doi
  [\mnras] {10.1093/mnras/stac2709}, \href
  {https://ui.adsabs.harvard.edu/abs/2022MNRAS.517..832I} {517, 832}

\bibitem[\protect\citeauthoryear{{Larson}}{{Larson}}{1976}]{Larson1976}
{Larson} R.~B.,  1976, \mn@doi [\mnras] {10.1093/mnras/176.1.31}, \href
  {https://ui.adsabs.harvard.edu/abs/1976MNRAS.176...31L} {176, 31}

\bibitem[\protect\citeauthoryear{{Marinacci}, {Pakmor}  \&
  {Springel}}{{Marinacci} et~al.}{2014}]{Marinacci2014}
{Marinacci} F.,  {Pakmor} R.,   {Springel} V.,  2014, \mn@doi [\mnras]
  {10.1093/mnras/stt2003}, \href
  {https://ui.adsabs.harvard.edu/abs/2014MNRAS.437.1750M} {437, 1750}

\bibitem[\protect\citeauthoryear{{Matteucci}}{{Matteucci}}{2012}]{Matteucci12}
{Matteucci} F.,  2012, {Chemical Evolution of Galaxies}.
Astronomy and Astrophysics Library, \mn@doi{10.1007/978-3-642-22491-1}

\bibitem[\protect\citeauthoryear{{Monachesi} et~al.,}{{Monachesi}
  et~al.}{2019}]{Monachesi2019}
{Monachesi} A.,  et~al., 2019, \mn@doi [\mnras] {10.1093/mnras/stz538}, \href
  {https://ui.adsabs.harvard.edu/abs/2019MNRAS.485.2589M} {485, 2589}

\bibitem[\protect\citeauthoryear{{Nuza}, {Scannapieco}, {Chiappini},
  {Junqueira}, {Minchev}  \& {Martig}}{{Nuza} et~al.}{2019}]{Nuza2019}
{Nuza} S.~E.,  {Scannapieco} C.,  {Chiappini} C.,  {Junqueira} T.~C.,
  {Minchev} I.,   {Martig} M.,  2019, \mn@doi [\mnras] {10.1093/mnras/sty2882},
  \href {https://ui.adsabs.harvard.edu/abs/2019MNRAS.482.3089N} {482, 3089}

\bibitem[\protect\citeauthoryear{{Okalidis}, {Grand}, {Yates}  \&
  {Kauffmann}}{{Okalidis} et~al.}{2021}]{Okalidis2021}
{Okalidis} P.,  {Grand} R. J.~J.,  {Yates} R.~M.,   {Kauffmann} G.,  2021,
  \mn@doi [\mnras] {10.1093/mnras/stab1142}, \href
  {https://ui.adsabs.harvard.edu/abs/2021MNRAS.504.4400O} {504, 4400}

\bibitem[\protect\citeauthoryear{{Okamoto}, {Eke}, {Frenk}  \&
  {Jenkins}}{{Okamoto} et~al.}{2005}]{Okamoto2005}
{Okamoto} T.,  {Eke} V.~R.,  {Frenk} C.~S.,   {Jenkins} A.,  2005, \mn@doi
  [\mnras] {10.1111/j.1365-2966.2005.09525.x}, \href
  {https://ui.adsabs.harvard.edu/abs/2005MNRAS.363.1299O} {363, 1299}

\bibitem[\protect\citeauthoryear{{Pakmor}, {Marinacci}  \& {Springel}}{{Pakmor}
  et~al.}{2014}]{Pakmor2014}
{Pakmor} R.,  {Marinacci} F.,   {Springel} V.,  2014, \mn@doi [\apjl]
  {10.1088/2041-8205/783/1/L20}, \href
  {https://ui.adsabs.harvard.edu/abs/2014ApJ...783L..20P} {783, L20}

\bibitem[\protect\citeauthoryear{{Pakmor} et~al.,}{{Pakmor}
  et~al.}{2017}]{Pakmor2017}
{Pakmor} R.,  et~al., 2017, \mn@doi [\mnras] {10.1093/mnras/stx1074}, \href
  {https://ui.adsabs.harvard.edu/abs/2017MNRAS.469.3185P} {469, 3185}

\bibitem[\protect\citeauthoryear{{Pakmor}, {Guillet}, {Pfrommer}, {G{\'o}mez},
  {Grand}, {Marinacci}, {Simpson}  \& {Springel}}{{Pakmor}
  et~al.}{2018}]{Pakmor2018}
{Pakmor} R.,  {Guillet} T.,  {Pfrommer} C.,  {G{\'o}mez} F.~A.,  {Grand} R.
  J.~J.,  {Marinacci} F.,  {Simpson} C.~M.,   {Springel} V.,  2018, \mn@doi
  [\mnras] {10.1093/mnras/sty2601}, \href
  {https://ui.adsabs.harvard.edu/abs/2018MNRAS.481.4410P} {481, 4410}

\bibitem[\protect\citeauthoryear{{Palla}, {Matteucci}, {Spitoni}, {Vincenzo}
  \& {Grisoni}}{{Palla} et~al.}{2020}]{Palla2020}
{Palla} M.,  {Matteucci} F.,  {Spitoni} E.,  {Vincenzo} F.,   {Grisoni} V.,
  2020, \mn@doi [\mnras] {10.1093/mnras/staa2437}, \href
  {https://ui.adsabs.harvard.edu/abs/2020MNRAS.498.1710P} {498, 1710}

\bibitem[\protect\citeauthoryear{{Pan}, {Li}, {Lin}, {Wang}, {Fan}  \&
  {Kong}}{{Pan} et~al.}{2015}]{Pan2015}
{Pan} Z.,  {Li} J.,  {Lin} W.,  {Wang} J.,  {Fan} L.,   {Kong} X.,  2015,
  \mn@doi [\apjl] {10.1088/2041-8205/804/2/L42}, \href
  {https://ui.adsabs.harvard.edu/abs/2015ApJ...804L..42P} {804, L42}

\bibitem[\protect\citeauthoryear{{Planck Collaboration} et~al.,}{{Planck
  Collaboration} et~al.}{2014}]{Planck2014}
{Planck Collaboration} et~al., 2014, \mn@doi [\aap]
  {10.1051/0004-6361/201323003}, \href
  {https://ui.adsabs.harvard.edu/abs/2014A&A...566A..54P} {566, A54}

\bibitem[\protect\citeauthoryear{{Prantzos} et~al.,}{{Prantzos}
  et~al.}{2023}]{Prantzos2023}
{Prantzos} N.,  et~al., 2023, \mn@doi [\mnras] {10.1093/mnras/stad1551}, \href
  {https://ui.adsabs.harvard.edu/abs/2023MNRAS.523.2126P} {523, 2126}

\bibitem[\protect\citeauthoryear{{Scannapieco}, {Tissera}, {White}  \&
  {Springel}}{{Scannapieco} et~al.}{2008}]{Scannapieco2008}
{Scannapieco} C.,  {Tissera} P.~B.,  {White} S. D.~M.,   {Springel} V.,  2008,
  \mn@doi [\mnras] {10.1111/j.1365-2966.2008.13678.x}, \href
  {https://ui.adsabs.harvard.edu/abs/2008MNRAS.389.1137S} {389, 1137}

\bibitem[\protect\citeauthoryear{{Scannapieco}, {White}, {Springel}  \&
  {Tissera}}{{Scannapieco} et~al.}{2009}]{Scannapieco2009}
{Scannapieco} C.,  {White} S. D.~M.,  {Springel} V.,   {Tissera} P.~B.,  2009,
  \mn@doi [\mnras] {10.1111/j.1365-2966.2009.14764.x}, \href
  {https://ui.adsabs.harvard.edu/abs/2009MNRAS.396..696S} {396, 696}

\bibitem[\protect\citeauthoryear{{Schaye} et~al.,}{{Schaye}
  et~al.}{2015}]{Schaye2015}
{Schaye} J.,  et~al., 2015, \mn@doi [\mnras] {10.1093/mnras/stu2058}, \href
  {https://ui.adsabs.harvard.edu/abs/2015MNRAS.446..521S} {446, 521}

\bibitem[\protect\citeauthoryear{{Sommer-Larsen}, {G{\"o}tz}  \&
  {Portinari}}{{Sommer-Larsen} et~al.}{2003}]{Sommer-Larsen2003}
{Sommer-Larsen} J.,  {G{\"o}tz} M.,   {Portinari} L.,  2003, \mn@doi [\apj]
  {10.1086/377685}, \href
  {https://ui.adsabs.harvard.edu/abs/2003ApJ...596...47S} {596, 47}

\bibitem[\protect\citeauthoryear{{Springel}}{{Springel}}{2010}]{Springel2010}
{Springel} V.,  2010, \mn@doi [\mnras] {10.1111/j.1365-2966.2009.15715.x},
  \href {https://ui.adsabs.harvard.edu/abs/2010MNRAS.401..791S} {401, 791}

\bibitem[\protect\citeauthoryear{{Springel} \& {Hernquist}}{{Springel} \&
  {Hernquist}}{2003}]{Springel2003}
{Springel} V.,  {Hernquist} L.,  2003, \mn@doi [\mnras]
  {10.1046/j.1365-8711.2003.06206.x}, \href
  {https://ui.adsabs.harvard.edu/abs/2003MNRAS.339..289S} {339, 289}

\bibitem[\protect\citeauthoryear{{Stinson}, {Brook}, {Macci{\`o}}, {Wadsley},
  {Quinn}  \& {Couchman}}{{Stinson} et~al.}{2013}]{Stinson2013}
{Stinson} G.~S.,  {Brook} C.,  {Macci{\`o}} A.~V.,  {Wadsley} J.,  {Quinn}
  T.~R.,   {Couchman} H.~M.~P.,  2013, \mn@doi [\mnras] {10.1093/mnras/sts028},
  \href {https://ui.adsabs.harvard.edu/abs/2013MNRAS.428..129S} {428, 129}

\bibitem[\protect\citeauthoryear{{Vogelsberger}, {Genel}, {Sijacki}, {Torrey},
  {Springel}  \& {Hernquist}}{{Vogelsberger} et~al.}{2013}]{Vogelsberger2013}
{Vogelsberger} M.,  {Genel} S.,  {Sijacki} D.,  {Torrey} P.,  {Springel} V.,
  {Hernquist} L.,  2013, \mn@doi [\mnras] {10.1093/mnras/stt1789}, \href
  {https://ui.adsabs.harvard.edu/abs/2013MNRAS.436.3031V} {436, 3031}

\bibitem[\protect\citeauthoryear{{Wang}, {Dutton}, {Stinson}, {Macci{\`o}},
  {Penzo}, {Kang}, {Keller}  \& {Wadsley}}{{Wang} et~al.}{2015}]{Wang2015}
{Wang} L.,  {Dutton} A.~A.,  {Stinson} G.~S.,  {Macci{\`o}} A.~V.,  {Penzo} C.,
   {Kang} X.,  {Keller} B.~W.,   {Wadsley} J.,  2015, \mn@doi [\mnras]
  {10.1093/mnras/stv1937}, \href
  {https://ui.adsabs.harvard.edu/abs/2015MNRAS.454...83W} {454, 83}

\bibitem[\protect\citeauthoryear{{Weinberger}, {Springel}  \&
  {Pakmor}}{{Weinberger} et~al.}{2020}]{Weinberger2020}
{Weinberger} R.,  {Springel} V.,   {Pakmor} R.,  2020, \mn@doi [\apjs]
  {10.3847/1538-4365/ab908c}, \href
  {https://ui.adsabs.harvard.edu/abs/2020ApJS..248...32W} {248, 32}

\makeatother
\end{thebibliography}




\appendix{}

\section{The inside-out parameter and its relation with galaxy properties}
\label{app:correlations}

\begin{figure*}
    \centering
    \includegraphics[draft=false]{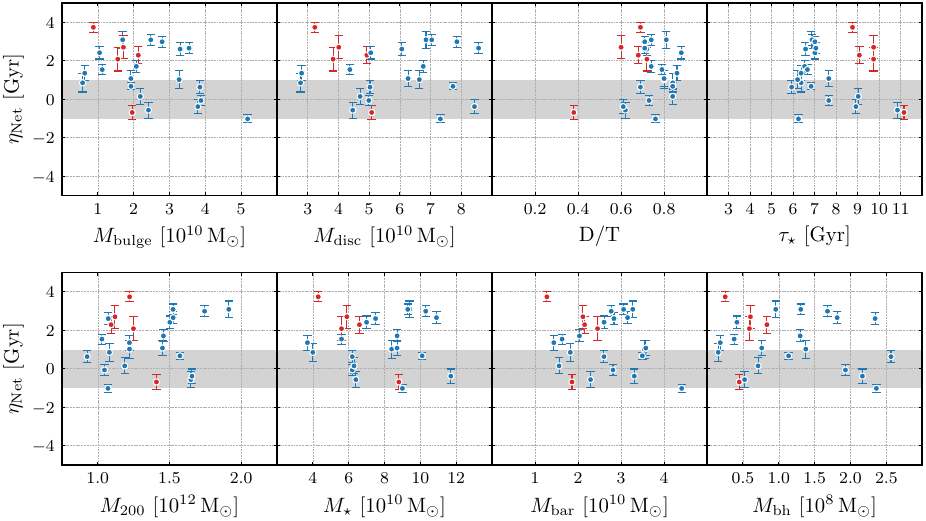}
    \label{fig:eta_vs_props}
    \caption{
    Correlation between the inside-out parameter $\eta$ calculated for the net accretion rates and the following galactic properties, from the top left to the bottom right: bulge mass $M_\mathrm{bulge}$ (calculated as the mass of stars inside $3~\mathrm{ckpc}$), disc stellar mass $M_\mathrm{disc}$, disc-to-total mass fraction D/T, stellar formation time $\tau_\star$, virial mass $M_{200}$, subhalo stellar mass $M_\star$, bar mass $M_\mathrm{bar}$, and black-hole mass $M_\mathrm{bh}$.
    All the values were calculated at $z=0$.
    The mass of the bar was calculated as $M_\mathrm{bar} = M_\star - 2 M_\star \left( \epsilon \geq 1 \right) - 2 M_\star \left( \epsilon \leq 0 \right)$, where $\epsilon$ is the circularity parameter defined in Paper I.
    To calculate $\tau_\star$ we use the time at which half the total stellar mass at $z=0$ is accumulated in the galaxy.
    }
\end{figure*}

In order the explore the possible relation of the inside-out growth of each simulated galaxy with its characteristics, we studied the correlation of the inside-out parameter with important global galaxy properties.
In Fig.~\ref{fig:eta_vs_props} we show the inside-out parameter, $\eta_{\rm Net}$, calculated using the net accretion rates (which is available for all simulations) as a function of the following galactic properties: (i) stellar mass of the bulge component, calculated as the total mass of stars inside a spherical region of size $3\,$kpc surrounding the galactic centre; (ii) stellar mass of the disc component, calculated as the mass of stars that lie inside the disc region defined in Paper I; (iii) disc-to-total mass ratio, calculated as described in Paper I; (iv) stellar formation time, calculated as the time at which half of the total stellar mass measured at $z=0$ is already in the galaxies; (v) total mass within a radius of $R_{200}$\footnote{$R_{200}$ is the radius inside which the mean mass density is $200$ times the critical density of the universe.}; (vi) stellar mass of the galaxy; (vii) mass of the galactic bar, and (viii) mass of the central black hole.
In each panel, we added a grey shade to the region $-1 \leq \eta_\mathrm{Net} \leq 1$, corresponding to galaxies inconsistent with an inside-out behaviour.

As discussed in the text, the different panels of Fig.~\ref{fig:eta_vs_props} show that there is no clear correlation between the inside-out parameter of each galaxy and their structural and dynamical properties at $z=0$.
This can not only be appreciated for the complete sample, but it is also valid if one isolates the haloes that display an inside-out behaviour (galaxies that lie above the shaded region in the figure).
Furthermore, the groups defined in Paper I are mixed when analysing the inside-out behaviour, meaning that the galaxies in each subsample, although they share common characteristics in terms of their formation history, show a variety of behaviours in terms of the inside-out parameter.
G1, for example, has samples both in the inside-out and the undefined regime (above and inside the shaded region, respectively).
G2 shows a similar distribution, but in this case the number of data points is considerably smaller.

\section{The inside-out parameter for different time intervals}
\label{app:dtt_inside_out}

Throughout this work, we quantified the inside-out behaviour of each of the Auriga galaxies based on the ``inside-out parameter'' $\eta$, defined in Eq.~\eqref{eq:tau}, and calculated between $t_{\rm i}=4~\mathrm{ Gyr}$ and the present day $t_0$.
However, it is possible to vary both $t_\mathrm{i}$ and $t_0$ to analyse the inside-out behaviour for specific time intervals.
In particular, for galaxies which do not follow an inside-out pattern during their whole evolution, we would like to study the disc-to-total mass ratio.

\begin{figure*}
    \centering
    \includegraphics[draft=false]{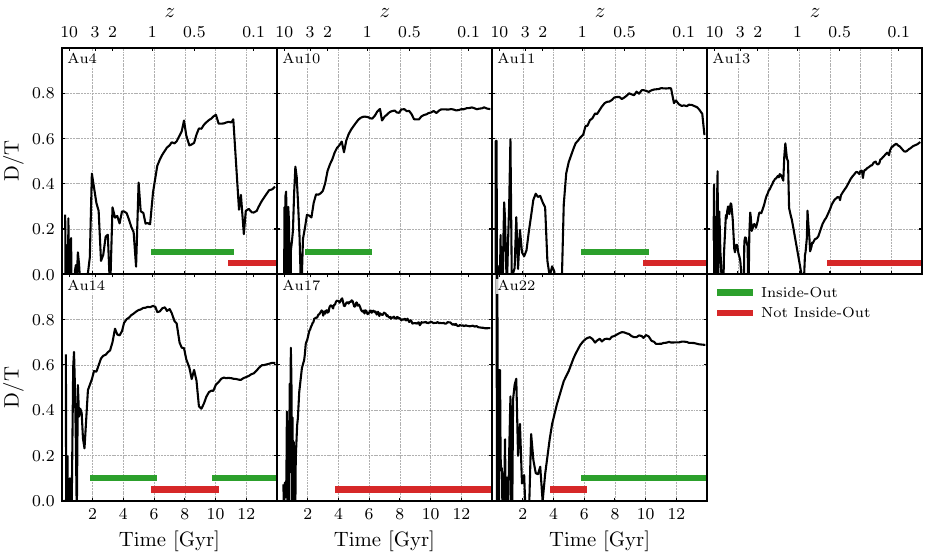}
    \caption{
        Evolution of the disc-to-total mass ratio for the galaxies that do not show a clear inside-out behaviour throughout their history: Au4, Au10, Au11, Au13, Au14, Au17, and Au22.
        The green (red) bars at the bottom of each panel indicate the time intervals where we obtain an inside-out (not inside-out) behaviour determined by computing the parameter $\eta$.
        For some of these galaxies, the inside-out periods coincide with times of disc growth, either after an initial chaotic formation stage (e.g., Au22) or after a (partial) destruction of the disc owing to a merger event (e.g., Au14).
    }
    \label{fig:dtt_with_inside_out}
\end{figure*}

In order to investigate this, we recalculated the $\eta$ values, for galaxies not classified as inside-out, for specific time periods of quiescent evolution and/or disc growth/destruction.
In Fig.~\ref{fig:dtt_with_inside_out} we show the evolution of the disc-to-total ratio through cosmic history for various of these galaxies, where we could define adequate, prolonged time intervals.
At the bottom of each panel, we show green and red bars indicating time periods where discs follow an inside-out (i.e. $\eta>0.8\,$Gyr) and an outside-in (i.e., $\eta<0.8\,$Gyr) pattern, respectively.
We find that, although these galaxies are not formally classified as inside-out according to our criterion, they show, in general, an inside-out behaviour during  periods in which the galactic discs are growing, as opposed to periods where discs are being destroyed or perturbed.
Au17 is, as explained in the text, the most extreme case and shows no sign of inside-out growth during the whole evolution.
A more detailed analysis of these galaxies is out of the scope of this paper and will be presented in future work.

\bsp	
\label{lastpage}
\end{document}